\documentclass{aa}
\usepackage{graphicx}
\begin{document}

%\refereelayout

\title{The role of absorption and reflection in the soft X-ray excess of Active Galactic Nuclei :\\
 1. Preliminary results}

\author{Lo\"{i}c Chevallier\inst{1}, Suzy Collin\inst{1}, Anne-Marie Dumont\inst{1}, Bozena Czerny\inst{2},\\
 Martine Mouchet\inst{1,3}, Anabela C. Gon\c{c}alves\inst{1,4}, Ren\'e Goosmann\inst{1}}

\date{09/10/2005}

\offprints{loic.chevallier@obspm.fr}

\institute{$^1$LUTH, Observatoire de Paris, Section de Meudon, F-92195 Meudon Cedex, France\\
$^2$Copernicus Astronomical Center, Bartycka 18, 00-716 Warsaw, Poland\\
$^3$APC, Universit\'e Denis Diderot, F-75005 Paris, France\\
$^4$Centro de Astronomia e Astrof\'{\i}sica da Universidade de Lisboa, Observat\'orio Astron\'omico de Lisboa, Tapada da Ajuda, 1349-018 Lisboa, Portugal}

\date{Received: 30/06/2005; Accepted: 21/10/2005}

\titlerunning{The role of absorption and reflection in AGN}
\authorrunning{L. Chevallier et al.}

\abstract{The 2-10 keV continuum of AGN is well represented by a single power law, generally attributed to a hot comptonizing medium, such as a corona above the accretion disk. At smaller energies the continuum displays an excess with respect to the extrapolation of this power law, called the ``soft X-ray excess". Until now it was attributed, either to reflection of the hard X-ray source by the accretion disk, or to the presence of an additional comptonizing medium. An alternative solution proposed by Gierli\'nski \& Done (2004) is that a single power law represents correctly both the soft and the hard X-ray emission, and the soft X-ray excess is an artefact due to the absorption of the primary power law by a relativistic wind. We examine the advantages and drawbacks of the reflection versus absorption models. We argue that in the absorption hypothesis, the absorbing medium should be in total pressure equilibrium, to constrain the spectral distribution which otherwise would be too strongly variable in time and  from one object to the other, as compared to observations. We  conclude that some X-ray spectra, in particular those with strong soft X-ray excesses, can be modelled  by absorption in the 0.3-10 keV range. However, due to the lack of a complete grid of models and good data extending above 10 keV, we are not able to conclude presently that all objects can be accommodated with such models. These absorption models imply either strong  relativistic outflowing winds with mass rates of the order of the Eddington value (or even larger), or quasi-spherical inhomogeneous accretion flows. Only weak excesses can be modelled by  reflection, unless the primary continuum is not directly seen. Finally, a reflection model absorbed by a modest relativistic wind could be the best solution to the problem.

}

\maketitle

\section{Introduction and rationale}
The X-ray spectra of radio quiet Active Galactic Nuclei (AGN) are well represented as one or more broad continuum components with superimposed spectral features due to absorption and emission by the circumnuclear material. The key problem is that the decomposition of the observed spectrum is not unique. 

The important role of both the emission and the absorption features in shaping the observed spectra is clear. Significant intrinsic absorption was already measured in Ariel 5 observation of NGC 4151 (Ives et al. 1976). The first X-ray narrow absorption feature was detected by Halpern (1984) in \textit{EINSTEIN} data. Now more than 50\% of well studied Seyfert galaxies and many quasars are known to possess absorbers at various ionization degrees (e.g. Seyfert 1: Nicastro et al. 2000, Behar et al. 2003, Steenbrugge et al. 2005, Young et al. 2005; Narrow Line Seyfert 1: Leighly et al. 1996, Gallo et al. 2004; PG QSO: Pounds et al. 2003a and 2003b, Porquet et al. 2004, Piconcelli et al. 2004, 2005; BAL QSO: Gallagher et al. 2002, Grupe et al. 2003). The first X-ray emission feature - the Fe fluorescent line - was already measured in 1978 (Mushotzky et al. 1978, Cen A). Later, the Compton reflection component was identified in a composite Ginga 12 spectrum (Pounds et al. 1990), and a broad Fe line was found (Tanaka et al. 1995) consistent with the expectations of a relativistically broadened feature due to X-ray reprocessing by an accretion disk. Nowadays many emission lines are identified in Seyfert 2 spectra (e.g. Sambruna et al. 2001, Kinkhabwala et al. 2002), iron lines are seen in most Seyfert 1 and 2 galaxies as well as in many quasars, and a few broad soft X-ray lines were possibly identified (Kaastra et al. 2002, R\' o\. za\' nska et al. 2004).  Relatively narrow features come from a significantly ionized medium, called the warm absorber (hereafter WA), which is outflowing with velocities from hundreds to thousands km s$^{-1}$ from the nuclear region (cf. for instance Blustin et al. 2005 for a collation and an analysis of the results  concerning 23 objects). This WA is located most probably somewhere between the Broad and the Narrow Line Region.

\begin{figure}
 \centering
 \includegraphics[width=9cm]{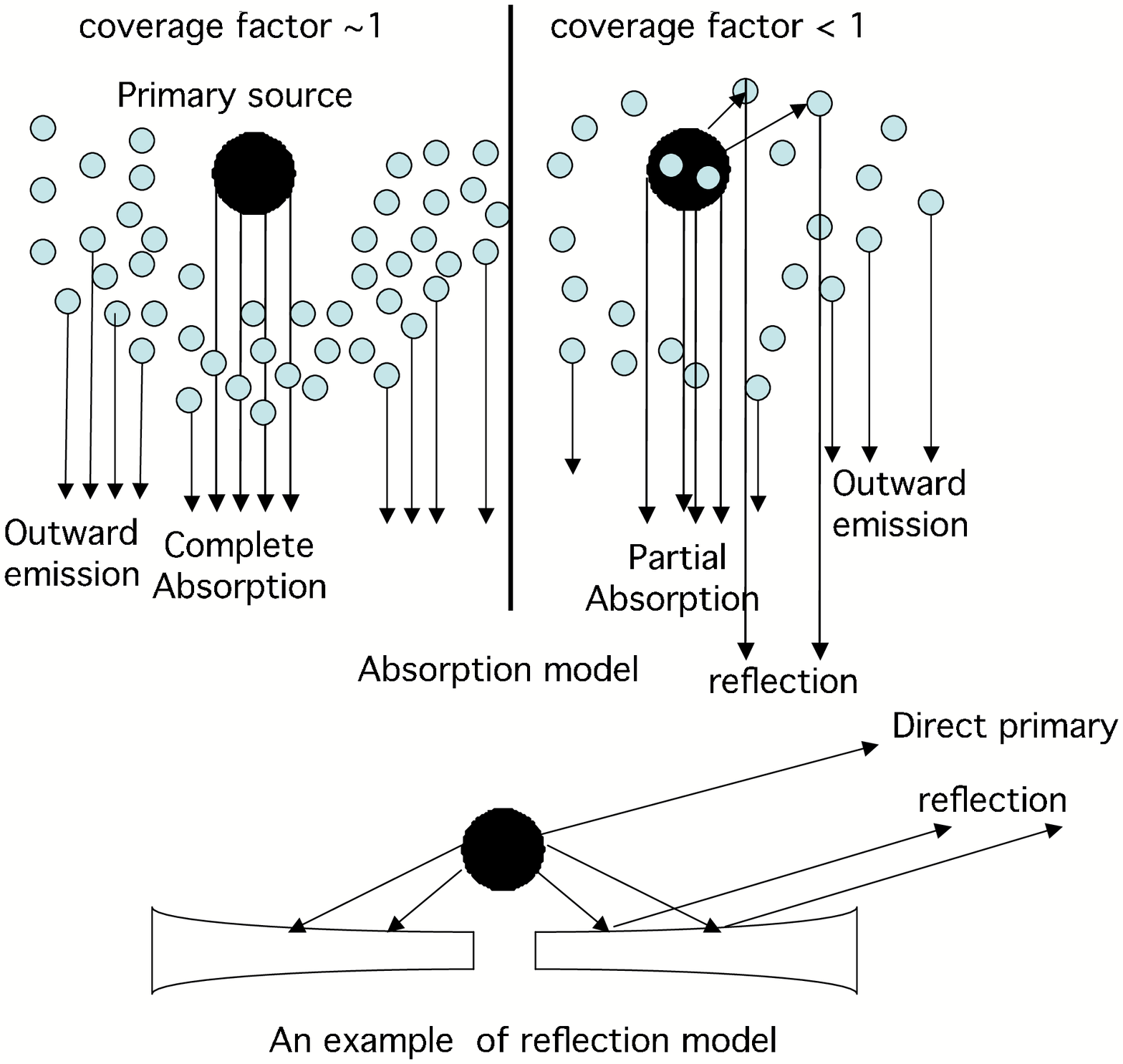}
\caption{Scheme of the absorption and reflection models showing the combination of spectra which constitutes the observed spectrum.}
 \label{fig-scheme}
\end{figure}

However, determining whether a given feature is due to absorption or emission is not always simple. There are two main issues which are still under discussion:

\noindent - The first topic is the question whether objects like Mkn~766 possess strong and relativistically smeared soft X-ray emission lines or the observed feature is actually due to the (dusty?) warm absorber (Branduardi-Raymont et al. 2001, Turner et al. 2003, Sako et al. 2003).

\noindent - The second topic is the question of the nature of the apparent slope change in the overall X-ray spectrum at $\sim$ 1 keV  in Seyfert 1 galaxies and radio quiet quasars. If the X-ray spectrum of an object is fitted with a power law plus absorption plus (eventually) the Compton reflection component plus (eventually) the iron line and (eventually) narrow spectral features, the model usually underpredicts the observed spectrum in the soft X-ray range. An additional component -- a soft X-ray excess -- is needed (Wilkes \& Elvis 1987).  The spectral shape and the nature of this component is under discussion since many years (e.g. Czerny \&  Zycki 1994). This component is usually modelled either as an additional continuum component (black body or multicolor blackbody; bremsstralhlung: Barvainis 1993; thermal Comptonization: Magdziarz et al. 1998), or as a strongly ionized reflection (i.e. scattering radiation with many atomic features like emission lines and recombination continua: Ross \& Fabian 1993, and subsequent works). However, as shown by Gierli\' nski \& Done (2004), this apparent change of slope is equally well modelled as being due to absorption of an originally rather soft power law intrinsic spectrum due to the warm absorber.

Resolving these issues is essential both for understanding the behaviour of the absorbing nuclear material and for the determination of the true intrinsic spectrum and, subsequently, for understanding the process of an accretion flow onto the central black hole.

It is natural that the presence of material surrounding the central source would affect the observed spectrum, and in particular lead to both emission and absorption features. The effect should depend on the location of the medium, its ionization, clumpiness, as well as the specific line of sight to the source. Some of the plasma parameters cannot be taken as arbitrary since the thermal state of the material is determined by the physical conditions. 

The aim of this paper is to consider the advantages and drawbacks of the reflection versus absorption models, by modeling the transfer of radiation in the plasma surrounding the central black hole, taking into account its natural physical limitations.
 
 In the next section, we recall some generalities about the models and our photoionization code. In Sect. 3 we consider pure absorption models, and we show that, unless a constraint of total pressure equilibrium is imposed, they should lead to a very large variation of the absorption spectrum. In Sect. 4 we consider intermediate models (partial coverage and absorption/emission), then reflection models, and we show that the observed spectrum is mainly a function of the coverage factor and of the column density. Finally, in the last section, we discuss some physical implications of these models.

\begin{figure}
 \centering
 \includegraphics[width=9cm]{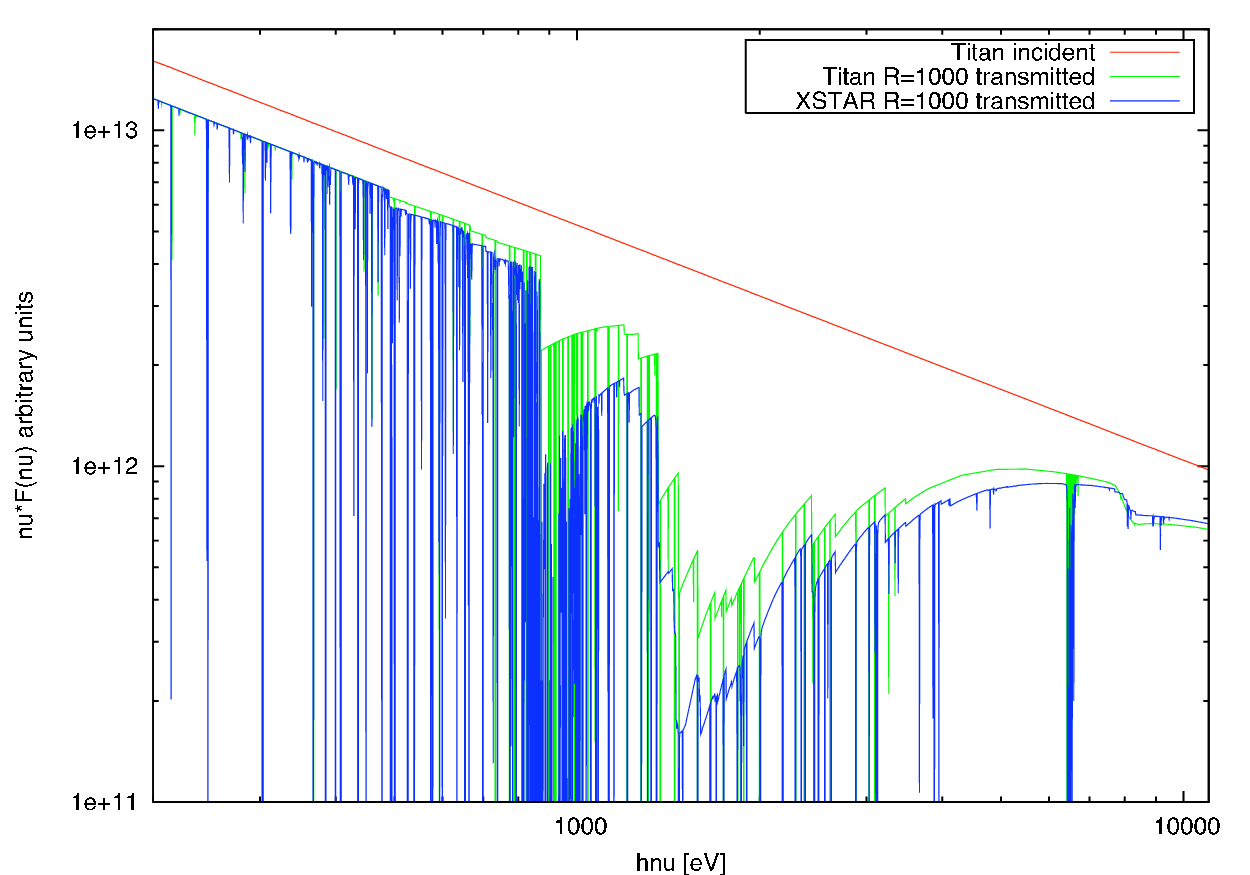}
 \includegraphics[width=9cm]{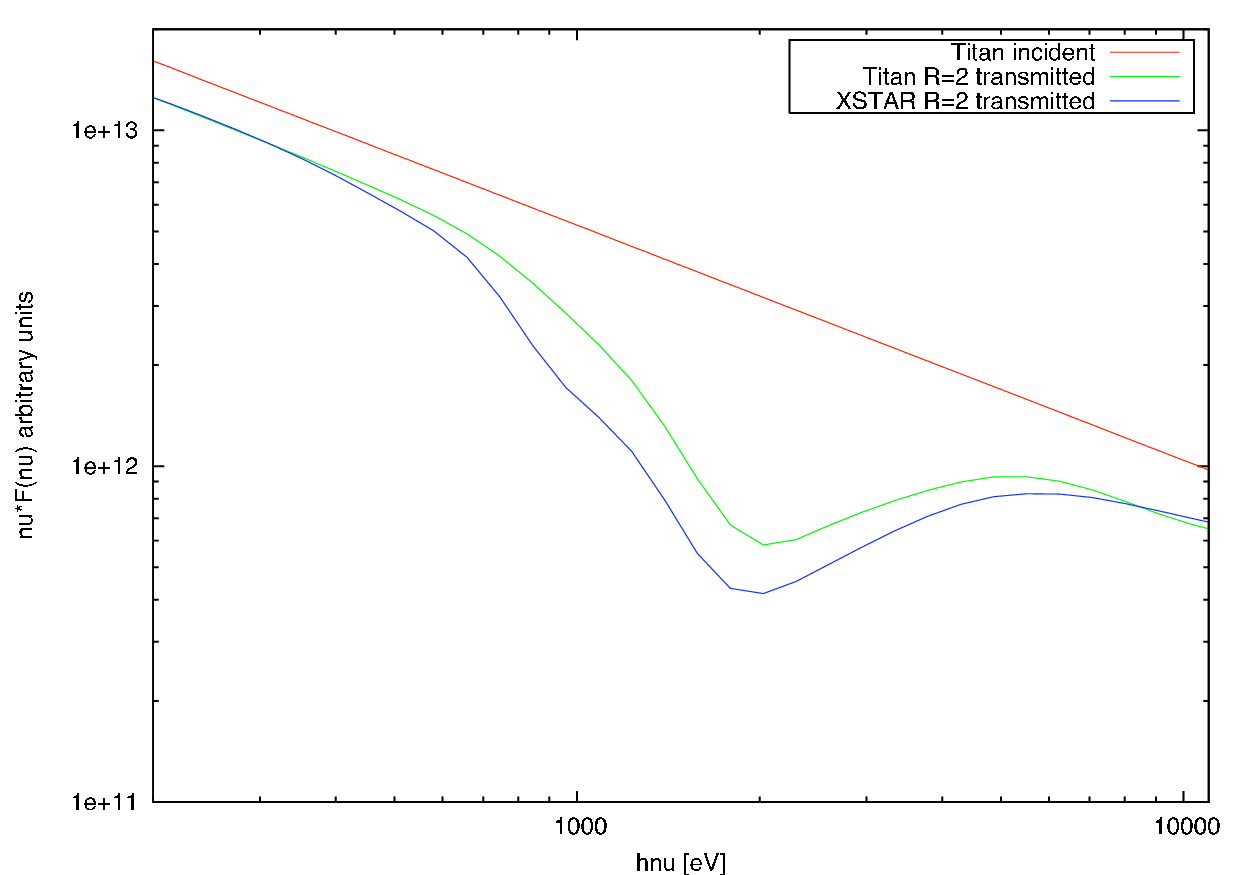}
\caption{Comparison of XSTAR and TITAN constant density calculations for physical conditions close to the Gierli\' nski \& Done (2004) best fitting model to PG 1211+143. Both panels represent the primary continuum and transmitted fluxes $\nu F_\nu$ (arbitrary units) vs. energy (in eV). The spectra are given at a spectral resolution 1000 (upper panel) -- differences for absorption lines, specially UTAs, are clearly visible -- and convolved with a velocity dispersion $v/c$=0.2 corresponding to a spectral resolution 2 (lowel panel). Differences on fluxes are less than 40 \%, and less than 10\% for the width of the trough taken at half maximum.}
 \label{fig-comp-GD}
\end{figure}

\begin{figure*}
 \centering
 \includegraphics[width=18cm]{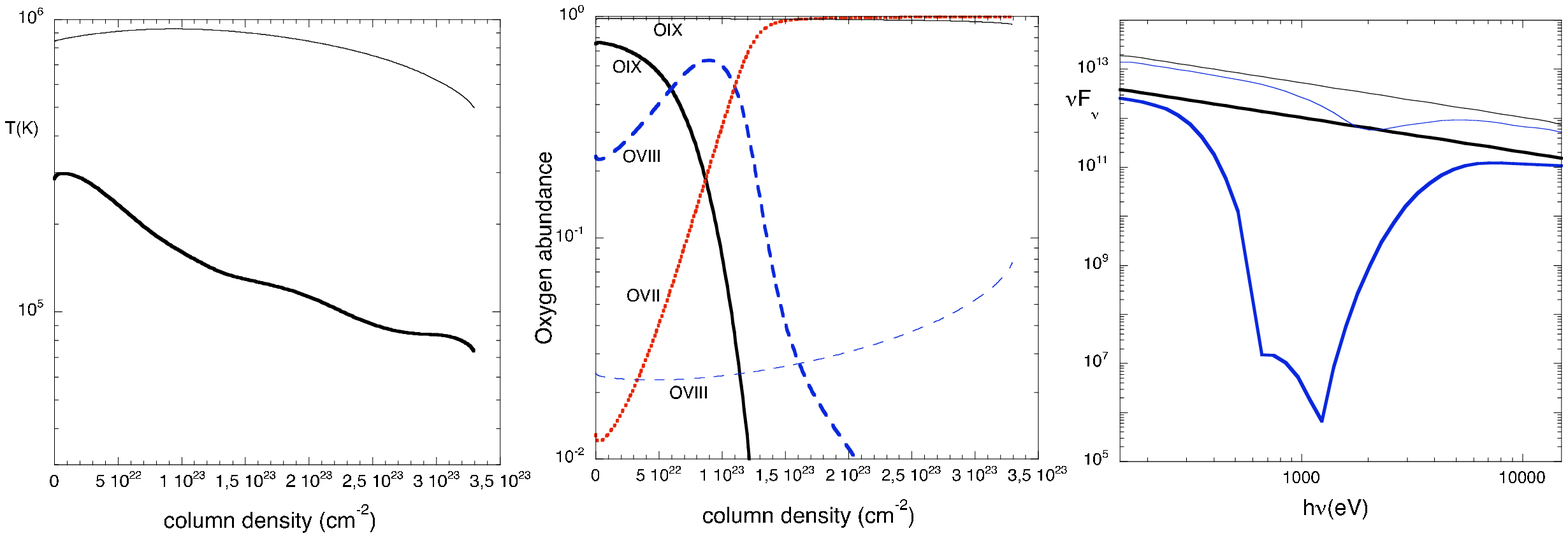}
\caption{Influence of the energy range of the primary continuum. The thin lines correspond to a primary continuum in the range 0.1 - 20 keV, and the thick lines show the corresponding results for a primary continuum in the range 0.01 - 100 keV. Both models have the same parameters as on Fig. \ref {fig-comp-GD} (note that the ionization parameter is integrated over the whole continuum, as explained in the text; this induces a slightly larger flux level for the ``truncated" continuum). For both models, we give the temperature profile (left panel), the fractional ionization abundances of oxygen (middle panel, including OVII to OIX for the ``extended'' continuum and OVIII and OIX -- which dominates -- for the ``truncated'' continuum) and the incident and absorption spectra (right panel). The spectral resolution is 2 and fluxes are in arbitrary units.}
 \label{fig-comp-bornes}
\end{figure*}

\section{Some generalities}
\label{sect:generalities}

As already mentioned, AGN are characterized by an X-ray spectrum extending into the gamma ray range up to hundred keV, or several hundreds keV. This emission is assumed to be provided by a region located close to the central black hole. Any  medium surrounding this primary source is radiatively heated and photoionized and will reprocess the primary photons as soft X-rays, UV, and possibly optical and IR photons. Whatever the structure of this medium  - clumpy or continuous - the observed spectrum will thus be a combination of ``reflection'' by the illuminated side (which is actually not a real reflection, as it includes atomic and Compton reprocessing), ``outward emission'' (emission by the back side), transmission of the incident primary spectrum, and possibly the primary spectrum itself. The amount of each component depends on the coverage factor of the source by the medium and on its geometry (see Fig. \ref{fig-scheme}).

If the irradiated medium has a flat and continuous structure, like an accretion disk, one will see a combination of about equal proportions of primary and reflected radiation. This is the most simple case, because if the medium is distributed quasi-spherically, all components  should be observed, though one or two of them can be negligible.  If its coverage factor is very close to unity, only the outward emission plus transmission will be observed. If the coverage factor is small, the primary source plus the reflection and the outward emission will be observed.

So {\it only in some cases will a single component be observed}: a pure absorption spectrum requires the source to be completely covered and a small column density, in order for the outward emission to be negligible. It is indeed difficult to imagine that the line of sight is completely covered, and that the other directions are not covered; it would mean that the absorbing medium is confined in a very small cone or filament along the line of sight. This is certainly neither the case of the WA which is observed in a large fraction of Seyfert 1s, nor the case of the Gierli\' nski \& Done (2004) model, which is assumed to account for the soft X-ray excess in all bright quasars. A pure reflection spectrum requires a peculiar disposition of the medium, as in the inhomogeneous accretion disk model proposed by Fabian et al. (2002, see the discussion below). 

All components are sensitive to several parameters. The thermal and ionization structure is determined mainly by the shape of the continuum in the soft X-ray range and by the ionization parameter. It is also not often realized that the upper and lower energies of the incident continuum play an important role in determining the free-free (at high density) and the Compton heating and cooling, as well as the ionization states of the elements. The absorption spectrum depends strongly on the column density, since the cool absorbing layers are located at the back side of our medium. The metal abundances influence not only directly the line spectrum but determine also the structure of the medium. On the contrary, micro-turbulence has a strong impact only on the line intensities in emission or in absorption, but a small influence on the thermal and ionization structure, at least for large column densities where the cooling is dominated by bound-free transitions. Also the density is not important, except for the relative intensities of the forbidden and permitted emission lines (for instance those of the helium-like ions like \ion{O}{vii}), and the overall spectrum is almost the same for a gas density varying from 10$^7$ to 10$^{12}$ cm$^{-3}$. It is why we do not specify the density used in the models, and why the  computed fluxes in the figures are always given in arbitrary units (they are proportional to the density).

	In this study, we used our photoionization code TITAN initially designed for ionized thick media (Thomson thickness up to several tens). The code is equally suited to model thinner media like the WAs. Its advantage over the other photoionisation codes like Cloudy (Ferland et al.  1998), XSTAR (Kallman \& Krolik 1995), or ION (Netzer 1993, 1996)  is that it treats the transfer of both the lines and the continuum using the powerful ALI method (Accelerated Lambda Iteration), which permits to compute very precisely line and continuum fluxes. The other photoionization codes use, at least for the lines, an integral formalism called  the ``escape probability approximation". In particular the computation of the absorption and emission lines is uncoupled when using this approximation, while with ALI both the lines and the continuum are treated in a consistent way.  In the context of X-ray spectra of AGN, the escape  approximation can lead to errors by a factor of several units on the line intensities. More important for the present study, in the case of thick media,  it leads to large errors in the temperature and ionisation structure near the back side of the medium which gives rise to the absorption features (Dumont et al. 2003, and Collin et al. 2004). 
	
	TITAN has been described in several papers (for instance Dumont et al. 2000, 2003), so we will not insist on its properties, recalling only that it solves the transfer of about 1000 lines and of the continuum, and gives as output the ionization and temperature structures, and the reflected, emitted outward, and absorbed spectra. The sophisticated line transfer treatment precludes presently to take into account as many lines as in Cloudy, XSTAR, or ION, so in particular for the WA it misses some important features (like the Unresolved Transition Array - UTA - around 750 eV, and the inner shell transitions, except the iron K lines). 
	
	The previous versions of our code used a 2-stream approximation for the transfer, which is equivalent to the ALI method with a 1-point angular quadrature (cosine=$1/\sqrt{3}$). This approximation is consistent with a semi-isotropic illumination. For the study of pure absorption spectrum corresponding to a normal illumination, it was necessary to perform first a computation with the 2-stream code to obtain the opacities and optical thicknesses of the model. Hence normal absorption was computed by the extinction of a normal incident flux for the same ionization parameter using the optical thicknesses obtained earlier (R\'o\.za\'nska et al. 2005). A recent  improvement was added to TITAN: to avoid this complex operation, we now use ALI in a multidirection version allowing to take into account a normal or inclined illumination. One can then determine the emitted or reflected intensity as a function of the direction. The advantage is to obtain the emission and absorption spectra in a completely consistent way for a  given opening angle and coverage factor of the medium.
	
	Our models consist in plane-parallel slabs illuminated on one side by an incident continuum. In the majority of the following models, the incident radiation field is concentrated in a small pencil normal to the slab. The value of the pencil opening angle is not important, provided it stays much smaller than a fraction of a radian. Otherwise the absorption spectrum is mixed with emission (cf. Sect.~4).
			
	Finally, a basic ingredient of the models is the ionization parameter $\xi$. Its definition varies among authors. We adopt $\xi = L/n_H R^2$ (in the following $\xi$ will be given always in erg cm s$^{-1}$), where $n_H$ is the
hydrogen number density at the illuminated surface, $R$ the distance between the primary continuum source and the photoionized medium, and $L$ is the luminosity of the continuum. In the following we integrate $L$ over the whole primary continuum (10 eV to 100 keV in this study), but some authors prefer to integrate the luminosity only over the 0.1-10 keV range where most of the X-ray absorption takes place, or from 0.54 to 10 keV,  the
energy range relevant for oxygen absorption, or from 1 to 1000 Rydberg as XSTAR does. The integration range should thus be taken into account when comparing the results of different authors.

\section{Pure absorption models}
\label{sect:absor}
  
\subsection{Constant density models}

In their paper, Gierli\' nski \& Done (2004) modelled the X-ray spectrum of the highly accreting source PG 1211+143 (a Narrow Line Seyfert 1, or NLS1) by a pure absorption spectrum produced by an ionized slab of constant density located on the line of sight of a source with a steep power law continuum between 0.1 and 20 keV. The photon index $\Gamma$ is equal to 2.7, quite a high value compared to typical observed indexes in the high energy range, but representative of radio quiet quasars and NLS1 in the soft X-ray band. The slab has a column density, $N$, equal to 3.3 $10^{23}$ cm$^{-2}$,  a turbulent velocity of 100 km s$^{-1}$, cosmic abundances, and $\xi = 460$  (these latter informations were kindly provided to us by C. Done). In order to get a ``quasi-continuum''
with no narrow features, Gierli\' nski \& Done assumed that the lines and the photoelectric edges are smeared by a large velocity dispersion $v$, so they convolved the absorbed spectrum with a Gaussian velocity dispersion $v/c= 0.2$ (FWHM = $2 \sqrt{2\ln{2}} \ v/c \sim 2.35 \ v/c$) corresponding to a spectral resolution $R = 1/(2.35\ v/c) \sim 2$ (definition of $R$ used throughout this study). This high velocity can be due to an accelerated outflow, or to a disk wind dominated by Keplerian motion; in the latter case it should be  produced very close to the black hole, at a distance around $(c/v)^2 = 25$ gravitational radii $R_{\rm G}$ ($R_{\rm G}=GM/c^2$). Such a model fits the data without any separate soft excess. Also Sobolewska \& Done (2004) have fitted the spectrum of  another Narrow Line Seyfert 1 (1H 0707-495) with an absorption model. 

We have built a set of models made of slabs of constant density, illuminated by power law continua, and we have computed the pure absorption spectra, assuming that there is no corresponding emission and that the slabs cover completely the primary source of radiation ($f_\mathrm{cov} \sim 1$).  We  illustrate the discussion first on Fig. \ref{fig-comp-GD} by a comparison between XSTAR and TITAN to get an idea of the influence of atomic data in our code, specially the missing UTAs. Using similar physical conditions as in Gierli\' nski \& Done (2004), with $\xi=2350$ for the incident flux from 10 eV to 100 keV, both codes provide similar results. The upper panel shows the incident and transmitted spectra for a high resolution ($\sim$ 1000), and the lower panel corresponds to those spectra being convolved with a dispersion velocity $v/c$ = 0.2 (i.e, spectral resolution 2). Since TITAN code take into account the electron scattering while this effect is neglected in XSTAR, the TITAN spectrum for the same incident flux is systematically below the XSTAR spectrum by a factor given by the Thomson optical depth of the medium, and the incident and transmitted fluxes at 200 eV are not equal in TITAN but equal in XSTAR. Therefore, for a better comparison of the effect of absorption on the spectra, we adjusted
the XSTAR spectrum to TITAN spectrum at 200 eV for the continuum. For these convolved spectra, the differences are weak (less than 40\%), and could be explained notably by the different approaches in the treatment of radiative transfer and energy balance. Also the influence of the atomic data, clearly visible at spectral resolution 1000 specially for the UTAs region around 900 eV, is weak for such a smearing. This comparison illustrates how much these models are sensitive to the computational methods, independently of the parameters (see also the comparison between several codes in Pequignot et al. 2001). We note however that for both models we see only one trough, located at around 2 keV, the absorption below 400 eV and above 8 keV are the same, and the difference for the width of this trough (taken at half maximum) is less than 10\%. UTAs included in XSTAR are the reason of a second trough around 800 eV, too weak to induce a difference in the absorption profile. \textit{Both codes give similar results for the constant density case, and our TITAN code can be used safely to explain WA features in this study in spite of its weaker number of lines treated}.

The influence of the physical parameters is illustrated on the following figures, which display a few examples of pure absorption spectra. In order to see easily the influence of the absorption on the soft X-ray excess, we have computed all models with simple power law incident continua ($F_{\nu} \propto \nu^{-\alpha}$).

\begin{figure*}
 \centering
 \includegraphics[width=18cm]{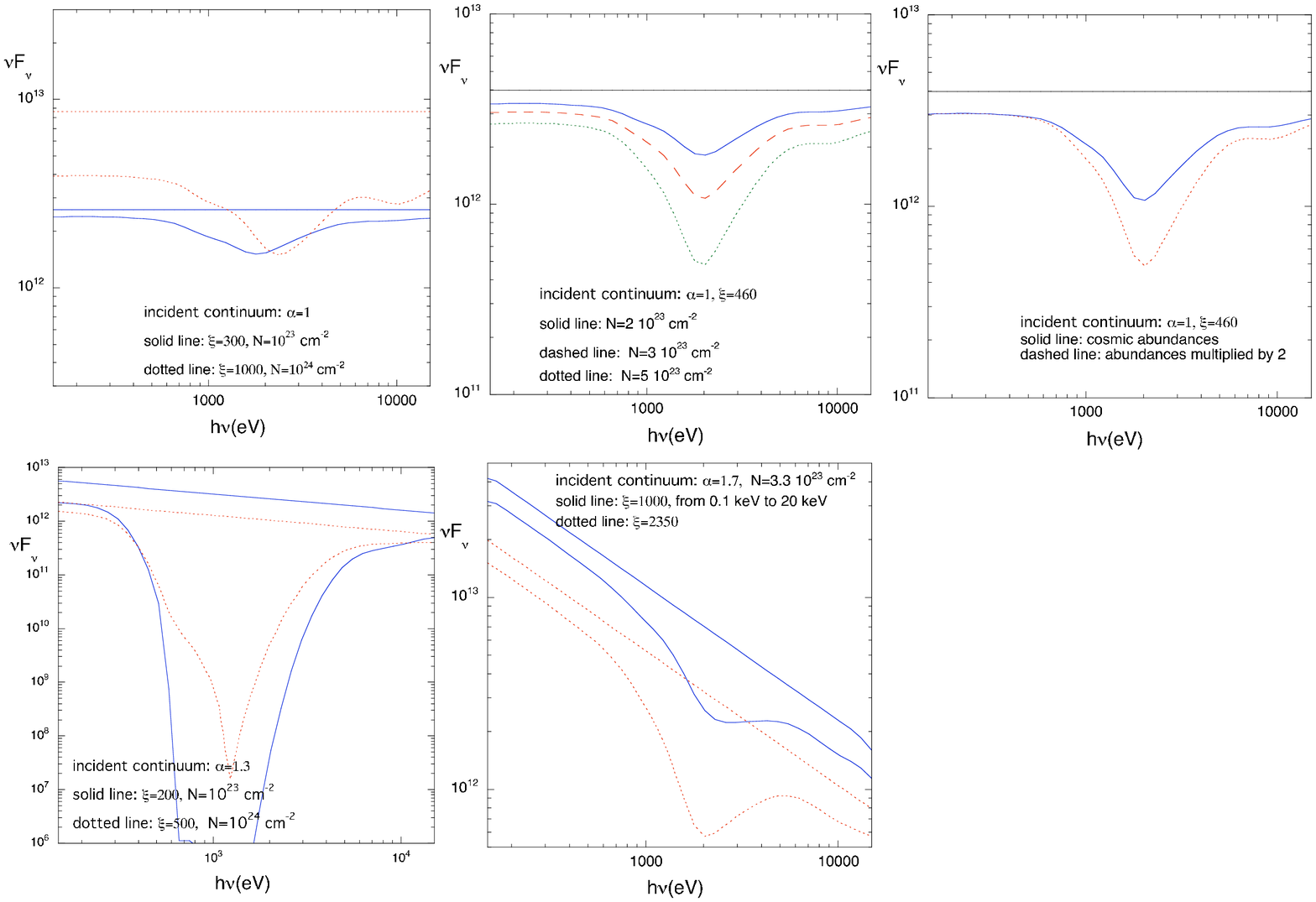}
\caption{Incident and absorption spectra for various constant density models, displayed with a spectral resolution of 2 for different models. The top panels correspond to a primary continuum with $\alpha=1$, and the panels on the bottom to continua with $\alpha$ larger than unity. The abundances are cosmic and the primary continua extend from 0.01 to 100 keV except when it is explicitely mentioned. Fluxes are in arbitrary units.}
 \label{fig-divers-spe1}
\end{figure*}

Figure \ref{fig-comp-bornes} illustrates the influence of the primary continuum energy range. The thin lines correspond to a primary continuum from 0.1 to 20 keV (actually the Gierli\' nski-Done model), and the thick lines show the corresponding results for a primary continuum covering the 0.01 - 100 keV range. Both models have the same density, same column density, same metal abundances, same microturbulent velocity,  and same ionization parameter (integrated over the whole continuum, which explains why the flux level is slightly larger for the ``truncated" continuum). The panel on the left shows the temperature profile: it is much higher for the ``truncated" continuum. Indeed with the ``extended" continuum the model includes more species able to cool the medium, as it is seen in the middle panel, which displays the fractional ionization abundances of oxygen for both models. For the ``truncated" continuum oxygen is mainly in the form of \ion{O}{IX}, while \ion{O}{VII} and \ion{O}{VIII} are predominant for the ``extended" continuum. As a result there is a strong imprint of the absorption  in the case of the ``extended" continuum, as it can be seen on the panel on the right where the spectra are displayed with a spectral resolution of 2.

Figure \ref{fig-divers-spe1} shows a few examples of absorption spectra. They are not aimed at illustrating in detail the influence of  the different physical parameters, but only at showing the variations from one spectrum to the other. All these models were calculated at a constant density. The top panels correspond to a primary continuum with $\alpha=1$, and the panels on the bottom to continua with $\alpha$ larger than unity. The abundances are cosmic except when it is mentioned. All primary continua extend from 0.01 to 100 keV except when it is mentioned (in the following we shall call ``standard" primary continuum the $\alpha=1$ power law within these limits). Among the top panels, the left one shows the influence of the ionization parameter, the middle one the influence of the column density, the right one the influence of the abundances. The bottom panels show the influence of the column density, the ionization parameter and the limits of the incident continuum. One can see that all absorption spectra have a trough around 1 keV. Its position is shifted towards slightly larger energies for higher values of the ionization parameter, as more ionized species become dominant. 

\subsection{A fine tuning prescription for pure absorption models: total pressure equilibrium}

A conclusion which can be drawn from Fig. \ref{fig-divers-spe1}  is that  the intensity of the trough varies strongly with the values of the parameters. Indeed several spectra of this figure would be completely incompatible with any observed X-ray spectrum, because the absorption is much too large, though all parameters vary in only relatively small ranges. As a consequence, any small variation with time of the column density on the line of sight, or of the flux of the primary continuum, would induce a strong variation on the shape of the X-ray spectrum.  It implies that there must be some kind of ``fine tuning" phenomenon which insures that the intensity of the trough cannot exceed some given value.

The previous models have been computed assuming, as it is generally done in this kind of problems, that the absorbing medium is one or several slabs of constant density. This is not necessarily appropriate, owing to the relatively short dynamical time scales of the absorbing gas. We will see in the last section that the pure absorption hypothesis implies a small distance between the primary source (or the black hole) and the absorber. The dynamics of the absorbing medium as a whole  would thus be dominated by the gravitation of the black hole, and its dynamical time $t_\mathrm{dyn}$ would  be of the order of the orbital time at the distance $R$ of the black hole: 

\begin{equation}
t_{\rm dyn} \sim {R\over v_{\rm abs}} \sim {1\over \Omega} =5\times 10^4 r_{10}^{3/2} M_7 \ \ {\rm s},
\label{tq-tdyn}
\end{equation}
where $v_{\rm abs}$ is a typical velocity of the absorber (rotation, outflowing velocity, large scale turbulence...), $\Omega$ is the angular velocity, $r_{10}$ the distance to the black hole expressed in 10$R_{\rm G}$, and $M_7$ is the mass of the black hole in units of 10$^7$ M$_{\odot}$. 

Note that the other important time scales (heating and cooling times, ionization and recombination times) are much smaller due to the high gas and radiation density in these central regions (for a discussion, see for instance Collin et al. 2003).  So if any perturbation of the pressure equilibrium is produced, another equilibrium state will be established in less than a day for $r_{10} \sim M_7 \sim 1$. It might not correspond to a constant pressure, actually.  If the dynamics of the absorbing medium is dominated by rotation, it could be close to the hydrostatic equilibrium of an accretion disk in the gravitation potential of the black hole. If it is dominated by outflowing motions it would be a dynamical equilibrium. 

As a working approximation, let us assume that the medium reaches a static pressure equilibrium state determined by the total pressure, taken as the sum of the radiation pressure and the thermal gas pressure.

\begin{figure*}
 \centering
 \includegraphics[width=15cm]{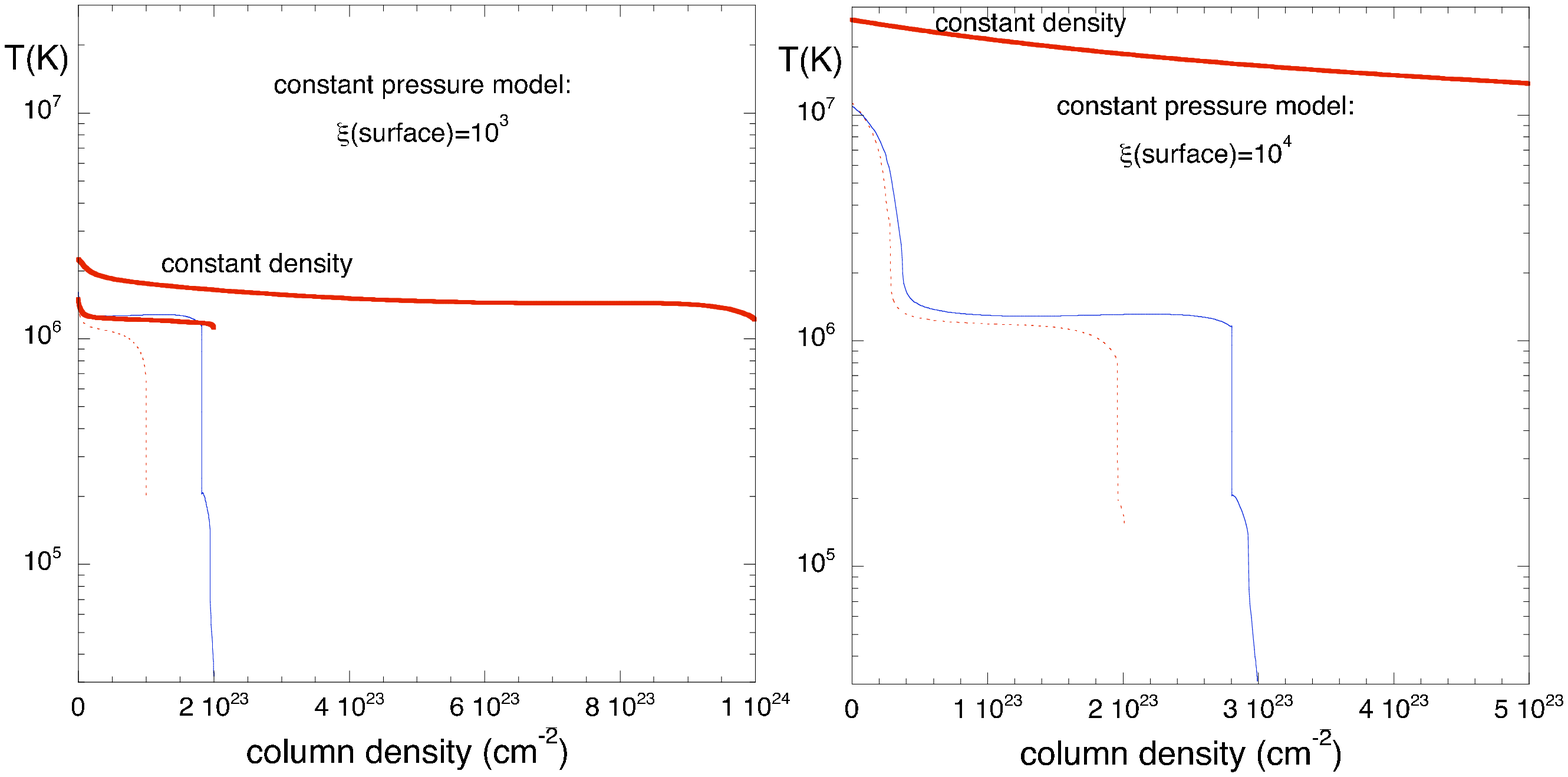}
\caption{Temperature profiles for several constant total pressure models (thin lines) computed with the standard primary continuum and two values of the ionization parameter at the surface, $\xi=10^3$ and $\xi=10^4$. The solid lines correspond to the maximum thickness, while the dotted lines correspond to models with a smaller thickness. The figures also show the temperature profile for constant density models  (thick lines) with the same values of the ionization parameter and of the surface density. }
 \label{fig-pcst-ncst-compCD-T}
\end{figure*}

\begin{figure*}
 \centering
 \includegraphics[width=15cm]{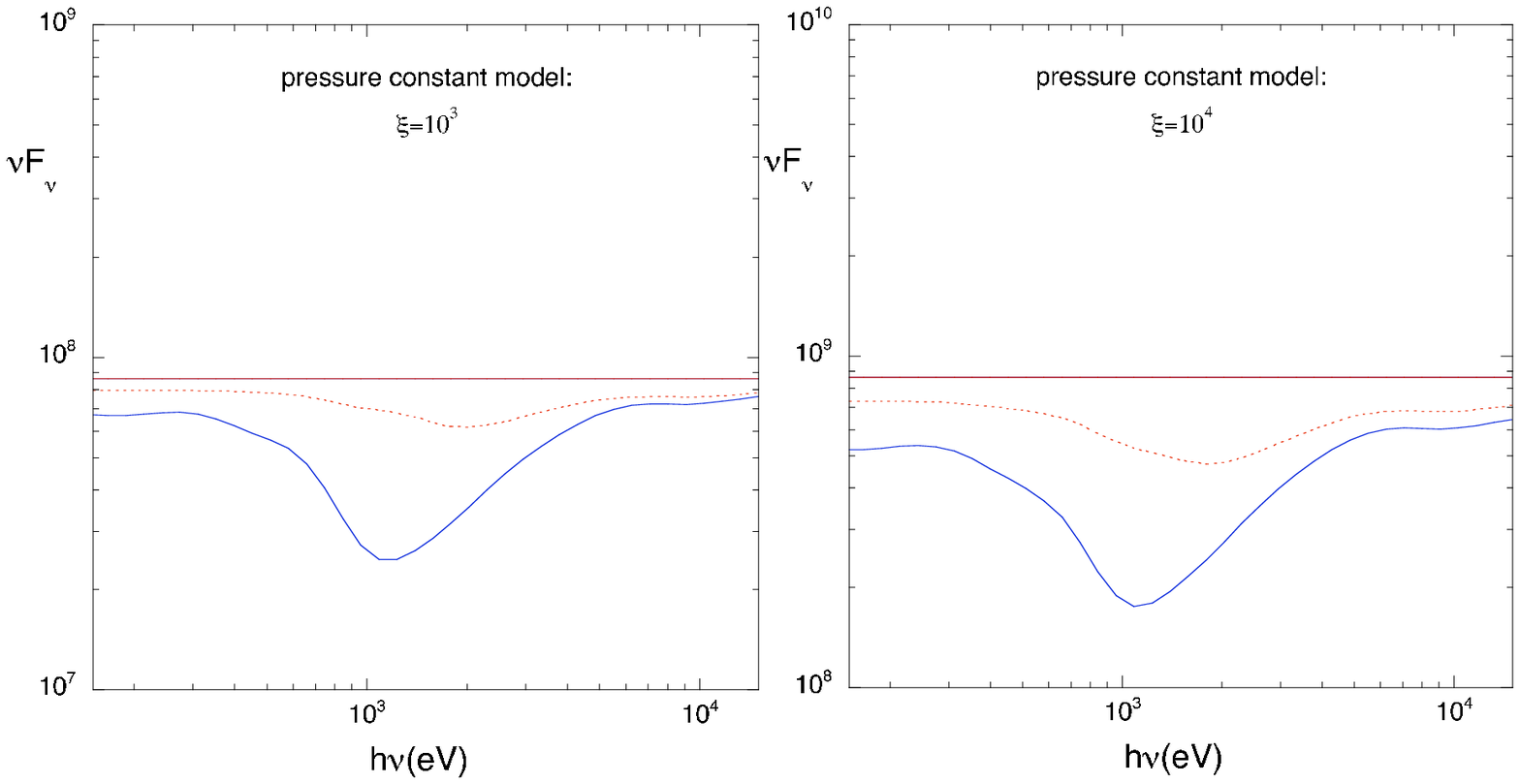}
\caption{Incident and absorption spectra corresponding to the constant total pressure models of Fig. \ref{fig-pcst-ncst-compCD-T}, displayed with a spectral resolution of 2. The solid lines correspond to the maximum thickness and the dotted lines to a smaller thickness. The straight line is the primary continuum. Fluxes are in arbitrary units.}
 \label{fig-pcst-compCD-spe1}
\end{figure*}

An important aspect of pressure - or hydrostatic - equilibrium models is that {\it the thickness of the illuminated slab cannot exceed a maximum value  for a given ionization parameter} (R\' o\. za\' nska et al. 2005). In a constant density slab, the temperature and the ionization states of the various elements decrease slowly with increasing depth, owing to the absorption of the incident continuum and the reemission of a diffuse soft X-ray spectrum. When the column density is very large (say larger than 10$^{24}$ cm$^{-2}$), the back side of the slab becomes eventually cold and neutral, and the whole X-ray spectrum is absorbed, allowing us to see only optical and IR photons and, at the other extreme of the spectrum, gamma-ray photons. The object would thus be ``Compton thick", and would require a different study. 

The behaviour of a slab in total pressure equilibrium is completely different, as the illumination by X-rays induces a thermal instability beyond a given layer in the slab. The phenomenon is due to the
S-shape of the temperature versus  the radiation to gas pressure ratio, allowing the existence of two or even more stable phases for the same gas pressure.  At a given gas pressure,  which depends on the energy distribution of the specific intensity,
the gas can be in three states of thermal equilibrium, corresponding to a hot and a cold stable solution, and to an unstable intermediate solution. Sometimes there are even five states with  three stable ones and two intermediate unstable ones.  This was shown by Krolik et al. (1981) for an optically thin gas. In our case the phenomenon is slightly different, as the shape of the radiation spectrum and the radiation pressure itself depend on the location in the medium. In the deepest layers, close to the back surface of the slab, the radiation spectrum becomes harder, and contains only hard X-rays, and as a consequence the S-shape curve is more pronounced, inducing an important instability and producing a very strong jump in temperature when the gas adjusts to the cold solution.   
 
  The thermal instability problem was already discussed in the context of the TITAN code and of hydrostatic equilibrium  by R\' o\.  za\' nska et al. (2002), so we will just recall it here. It is difficult to know in which state (hot or cold) the gas can be when both are allowed.  It probably depends on the previous history of the medium. Taking into account conductivity effects can help to solve the problem (R\' o\. za\' nska \& Czerny 2000), but it is a very difficult task when conductivity should be coupled with a complete transfer treatment like that of TITAN. 
  
  The present scheme of TITAN keeps the density constant in a given layer while searching for the equilibrium temperature. Such a numerical scheme produces a unique but approximate solution (intermediate between the hot and the cold solutions) in the unstable region.  We are thus developing a new algorithm allowing to choose between the hot and the cold exact stable solutions. Even with the present scheme the approximate computation is possible only when the thermal instability is not too much pronounced. When the temperature drops suddenly  to very  low values (of the order of $10^4$ K), the radiation pressure becomes dominated by spectral lines and induces a numerical instability which corresponds to a real thermal instability. Thus the slab should necessarily be broken into cold dense clumps, possibly embedded into a warm dilute medium, and it does not exist any more as an entity. However, like in the constant density case, a cold region completely absorbing the primary source can exist beyond this instability. 
  
   A consequence of this maximum thickness of the slab is the existence of a ``maximum absorption trough", which cannot be exceeded. Smaller values of the trough can be obtained if the column density is smaller. In this case there is also a decrease of temperature at the back of the slab associated with an absorption, but the drop of temperature is smaller. The decrease occurs also in constant density models near the back surface, as the radiation can escape more easily and the medium cools. Simply it is more pronounced in constant total pressure models because as the radiation pressure decreases it induces an increase in the gas pressure and therefore in the density, so the medium cools even more rapidly. All these questions will be addressed in more detail in a future paper. 

This discussion is illustrated on Fig. \ref{fig-pcst-ncst-compCD-T} which shows the temperature profile for several constant total pressure models (the thin lines). They have been computed with our standard primary continuum. The left panel corresponds to an ionization parameter at the surface $\xi=10^3$ and the right one to $\xi=10^4$. The solid lines correspond to the maximum thickness, and the dotted lines to models where the thickness has been imposed. In all cases the temperature decreases abruptly close to the back side, and has almost the same profile for the two ionization parameters. For comparison the two panels show also the temperature profiles for constant density models  (the thick lines), with the same values of the ionization parameter and of the surface density. One sees that {\it the temperature is almost constant and always high when the density is constant}. We have added a constant density model with a very large column density (10$^{24}$ cm$^{-2}$), only to show that the temperature stays high for a  larger thickness than the constant total pressure models. Note also that the thicker the medium, the higher the temperature, owing to the increased  heating by radiation ``returning" from the back (cf. Dumont et al. 2000). 

\begin{figure}
 \centering
 \includegraphics[width=7cm]{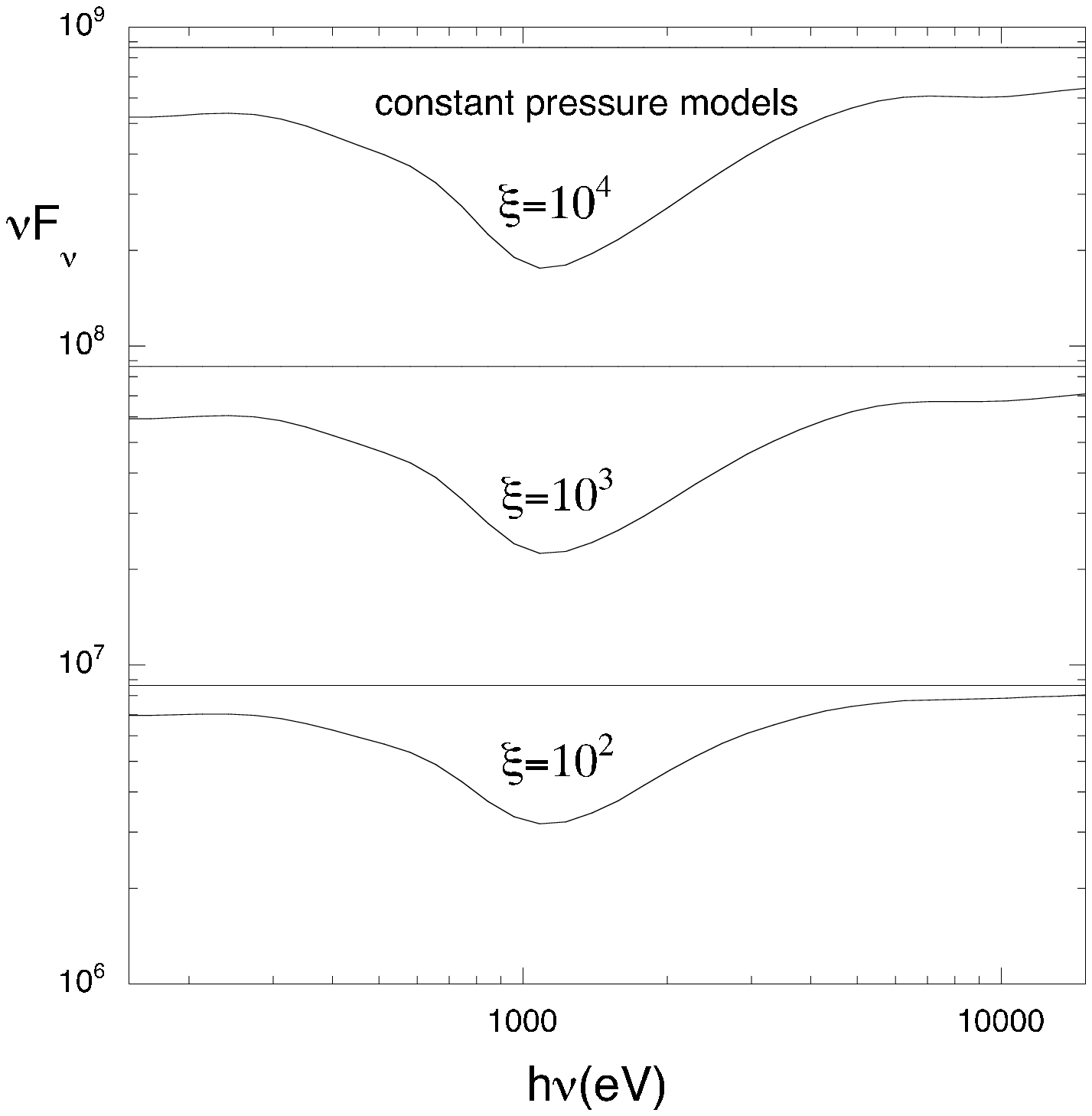}
\caption{ Absorption spectra for 3 constant total pressure models which have the maximum possible thickness for their given ionization parameter. The straight lines represent the primary continuum. The spectra are displayed with a spectral resolution of 2. Fluxes are in arbitrary units.}
 \label{fig-pcst-comp-spe1}
\end{figure}

\begin{figure*}
 \centering
 \includegraphics[width=18cm]{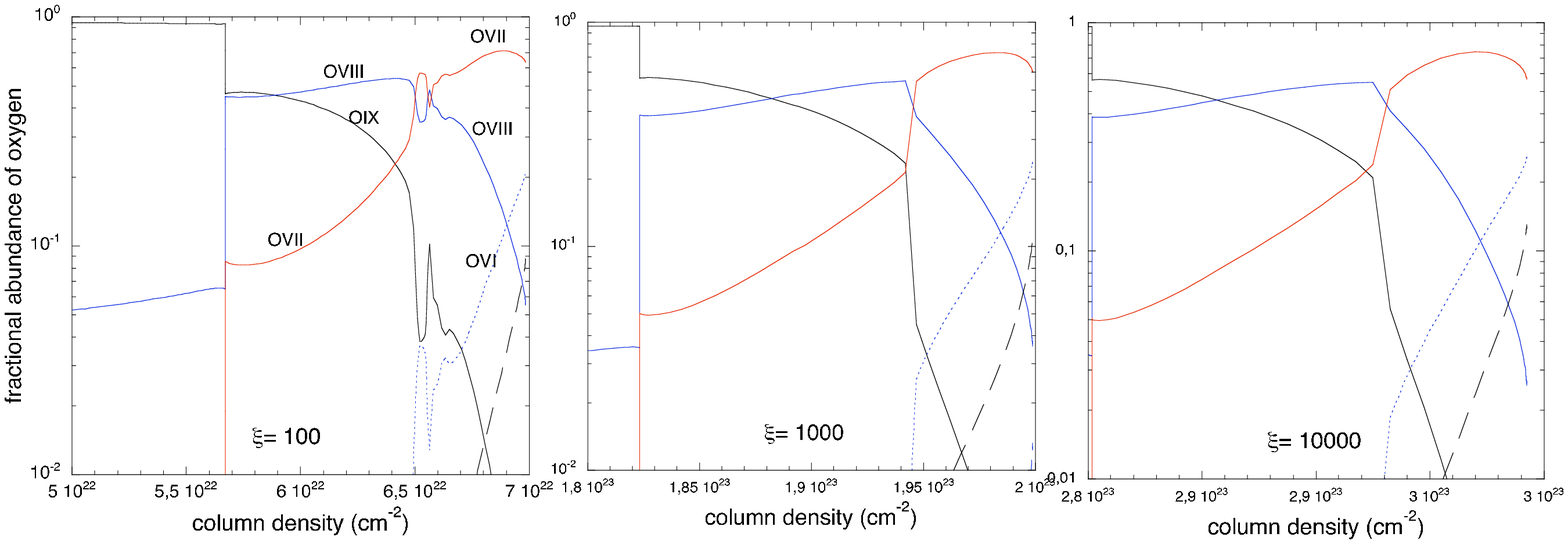}
\caption{Fractional ionic abundances of oxygen near the back surface, as a function of the column density in the slab, for the three models of Fig. \ref{fig-pcst-comp-spe1}. The abscissae range corresponds to exactly the same total thickness of 10$^{22}$ cm$^{-2}$. One can see that the fractional abundances vary very similarly, though the range of $\xi$ is quite large, and consequently the OVII and OVIII layers are located at very different distances from the illuminated sides.}
 \label{fig-pcst-deg0}
\end{figure*}

Figure \ref{fig-pcst-compCD-spe1} displays the incident and absorption spectra for the same constant total pressure models. The solid lines correspond to the maximum thickness and the dotted lines to a smaller thickness. Though the shape of the trough depends on the thickness of the slab, we see that {\it its strength cannot exceed a given value which does not seem to depend on the ionization parameter}. Considering only {\it constant pressure models with maximum thickness}, one can see that the shapes of the absorption spectra are very similar (cf. Fig. \ref{fig-pcst-comp-spe1}). In particular the absorption trough is located at the same energy, around 1 keV. Note however that the absorption {\it with respect to the primary continuum} is more important at all frequencies for a larger ionization parameter, and consequently a larger column density; this  is because the overall spectrum goes down due to the loss of photons through electron scattering. Unfortunately this effect cannot be directly measurable since the primary continuum is not observed.

It is easy to understand the similarity of the absorption spectra. One can see on Fig. \ref{fig-pcst-ncst-compCD-T}
 the temperature profiles of constant total pressure slabs with the maximum thickness, for $\xi$ = 10$^3$ and 10$^4$. It is really impressive how similar the temperature profiles are near the back surface. This is due to the fact that the gas enters in a multiple phase regime when the radiation spectrum has a given shape, whatever the physical state of the previous layers. This is well illustrated by Fig. \ref{fig-pcst-deg0} which shows the fractional ionic abundances of oxygen near the back surface, for the constant pressure models corresponding to the maximum thickness, for $\xi$ = 10$^2$, 10$^3$ and 10$^4$, as a function of the column density in the slab.  The abscissae range corresponds exactly to the same total thickness of  2 10$^{22}$ cm$^{-2}$.  One can see that the fractional abundances vary similarly, though the range of $\xi$ is quite large. This region is dominated by \ion{O}{vii} and \ion{O}{viii} ions which contribute to a large fraction of the absorption around 1 keV. The same result is obtained for the other elements, which explains why the absorption spectra are almost identical (cf. Fig. \ref{fig-pcst-compCD-spe1}). Therefore {\it the most important difference between constant density and constant total pressure models lies in the fact that the ``intermediate temperature layer" containing highly absorbing species, has the same thickness whatever its ionization parameter at the illuminated surface}. 

\begin{figure}
 \centering
 \includegraphics[width=9cm]{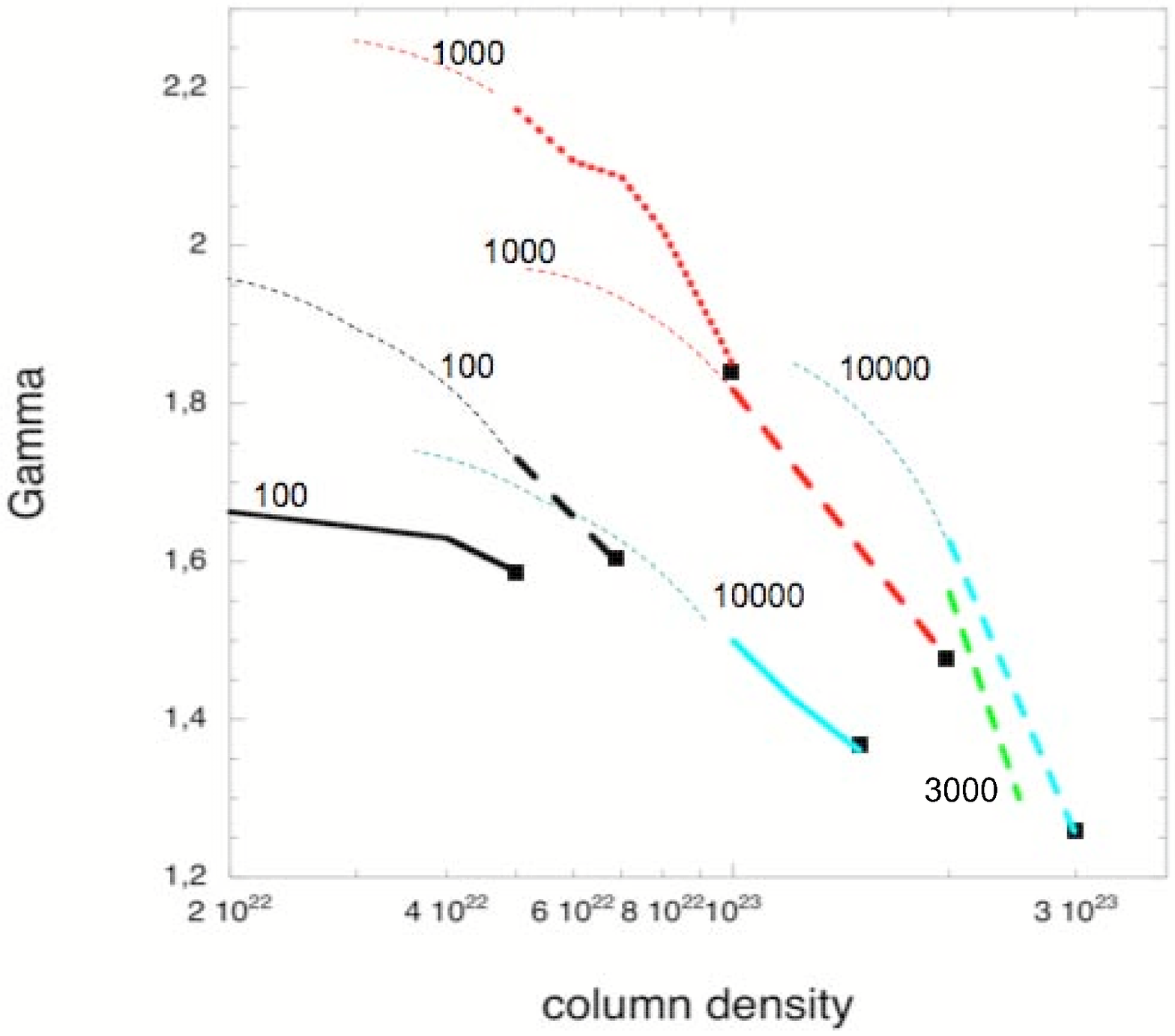}
\caption{ Value of  $\Gamma_{2-10 {\rm keV}}$ measured on the computed absorption spectra for constant total pressure models, as a function of the column density, for 3 values of the spectral index of the primary continuum. The curves are labelled with the value of the ionization parameter. Solid lines: $\alpha=0.8$, dashed lines: $\alpha=1$, dotted lines: $\alpha=1.3$. The squares mark the position of the maximum column density corresponding to a given ionization parameter. The small dashed lines are extrapolations which take into account that the upper value of $\Gamma$ for optically thin slabs is equal to $\alpha+1$. }
 \label{fig-modeles-gammas}
\end{figure}

\begin{figure}
 \centering
 \includegraphics[width=8cm]{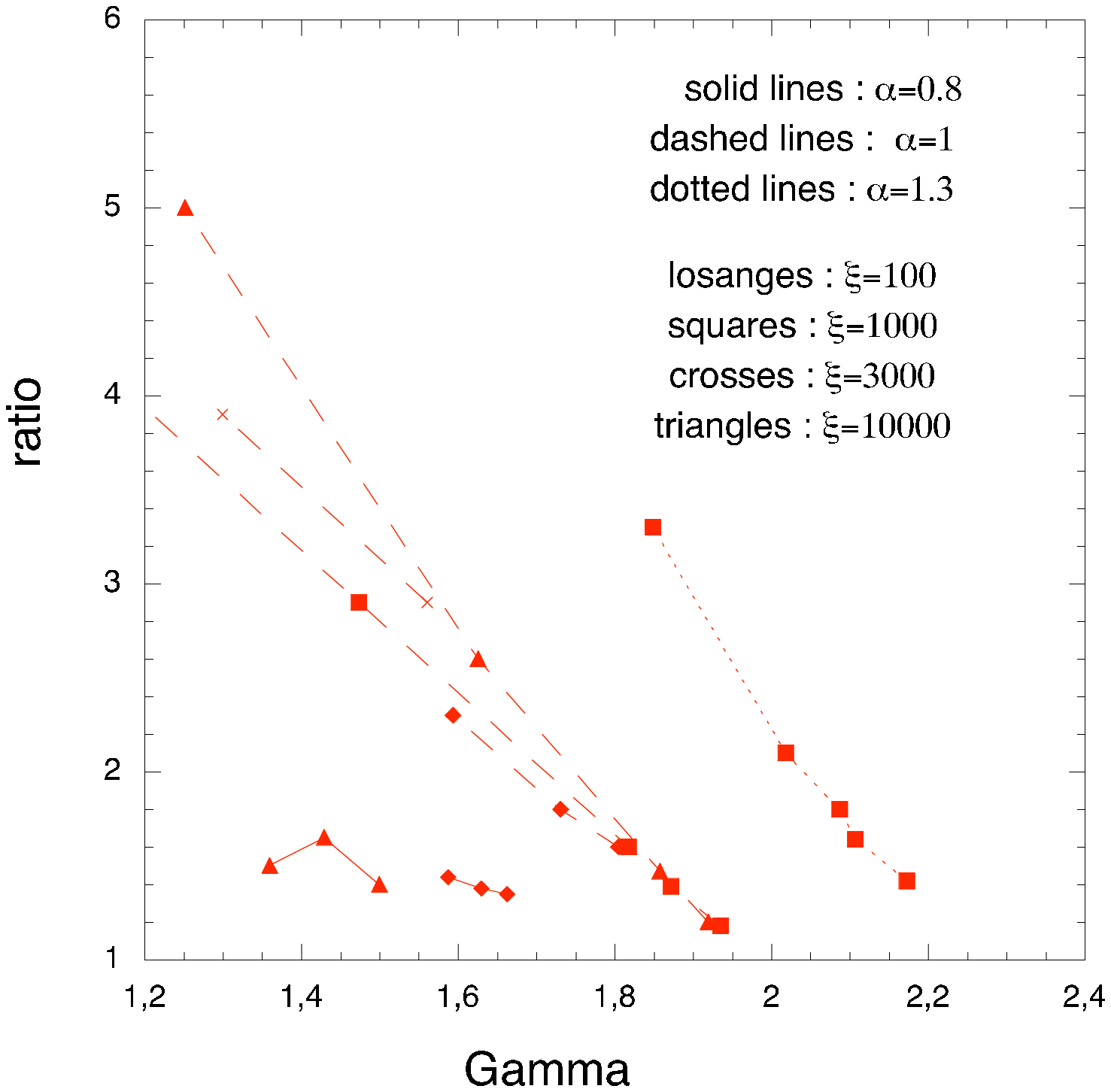}
\caption{ Ratio of the 0.5 keV flux to the extrapolation of the $\Gamma_{2-10 {\rm keV}}$ power law measured on the computed absorption spectra versus $\Gamma_{2-10 {\rm keV}}$, for different values of the spectral index of the primary continuum $\alpha$ and different values of $\xi$. }
 \label{fig-ratios-vs-gamma}
\end{figure}

\bigskip

With the computed absorption spectra, one can determine some parameters easily deduced from the observed spectra. We have chosen: 1) an ``average" photon spectral index $\Gamma_{2-10 {\rm keV}}$ measured from 2 to 10 keV and 2) the ratio of the 0.5 keV  flux to the extrapolation of this power law ($Ratio$). 

For different constant pressure models, Fig. \ref{fig-modeles-gammas} shows the variation of $\Gamma_{2-10 {\rm keV}}$ with the column density, for several values of the primary continuum spectral index $\alpha$ and of the ionization parameter at the illuminated surface $\xi$. Since the value of $\Gamma_{2-10 {\rm keV}}$ tends  towards that of the primary continuum $\alpha\ ({\rm primary\; continuum})+1$ for very thin slabs (because the absorption spectrum is identical to the primary one), we have extrapolated the curves taking this constraint into account. Figure \ref{fig-ratios-vs-gamma} displays several curves giving $Ratio$ versus $\Gamma$: they correspond to different values of $\alpha$ and of $\xi$. It is interesting to see that the main variable is $\alpha$ and that the curves corresponding to different values of $\xi$ are almost aligned.  It means that it would not be possible to find a unique fit for an observed spectrum, on the basis of these two parameters only.

Nevertheless, helped by the curves of {Fig. \ref{fig-modeles-gammas},  we have tried to fit some observed spectra with constant total pressure models.
Figure \ref {fig-comp-mod-obs} shows a comparison on PG 1307+085. The observed spectrum is the result of the ``best fit" of the EPIC data of {\it XMM}  with phenomenological models including blackbodies and power laws, obtained by Piconcelli et al. (2005, see their Fig. 2). We measure $\Gamma_{2-10 {\rm keV}} = 1.5$ for this object. For the sake of clarity we divided the spectra by a power law with this value of $\Gamma=1.5$. The model is a constant total pressure slab illuminated by a  power law continuum with $\alpha=0.9$, $\xi=10^4$, and $N$= 2 10$^{23}$ cm$^{-2}$. The absorption spectrum has been divided by the same power law as the observed spectrum and they are displayed with a spectral resolution of 2 (we recall that it corresponds to a dispersion velocity equal to $c/5$) and 100. This spectrum is  well fitted, considering that the narrow emission feature around 0.5 keV - the OVII complex - must be provided by another emitting region. Note that this model corresponds to a thickness smaller than the maximum value allowed for its ionization parameter.

\begin{figure}
 \centering
 \begin{tabular}{cc}
 \hspace{-0.5cm} & \includegraphics[width=8.5cm]{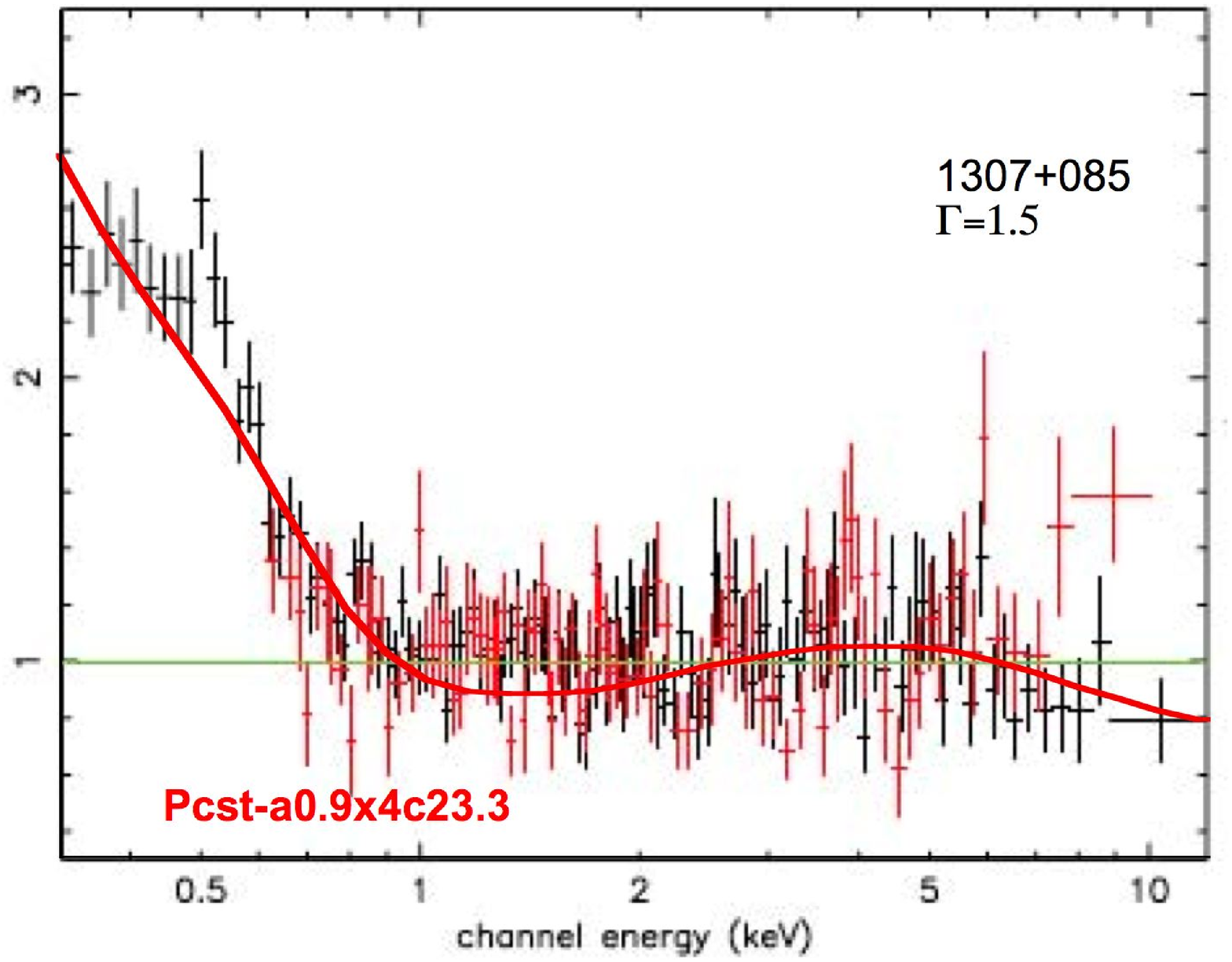}
 \end{tabular}
\includegraphics[width=8.3cm]{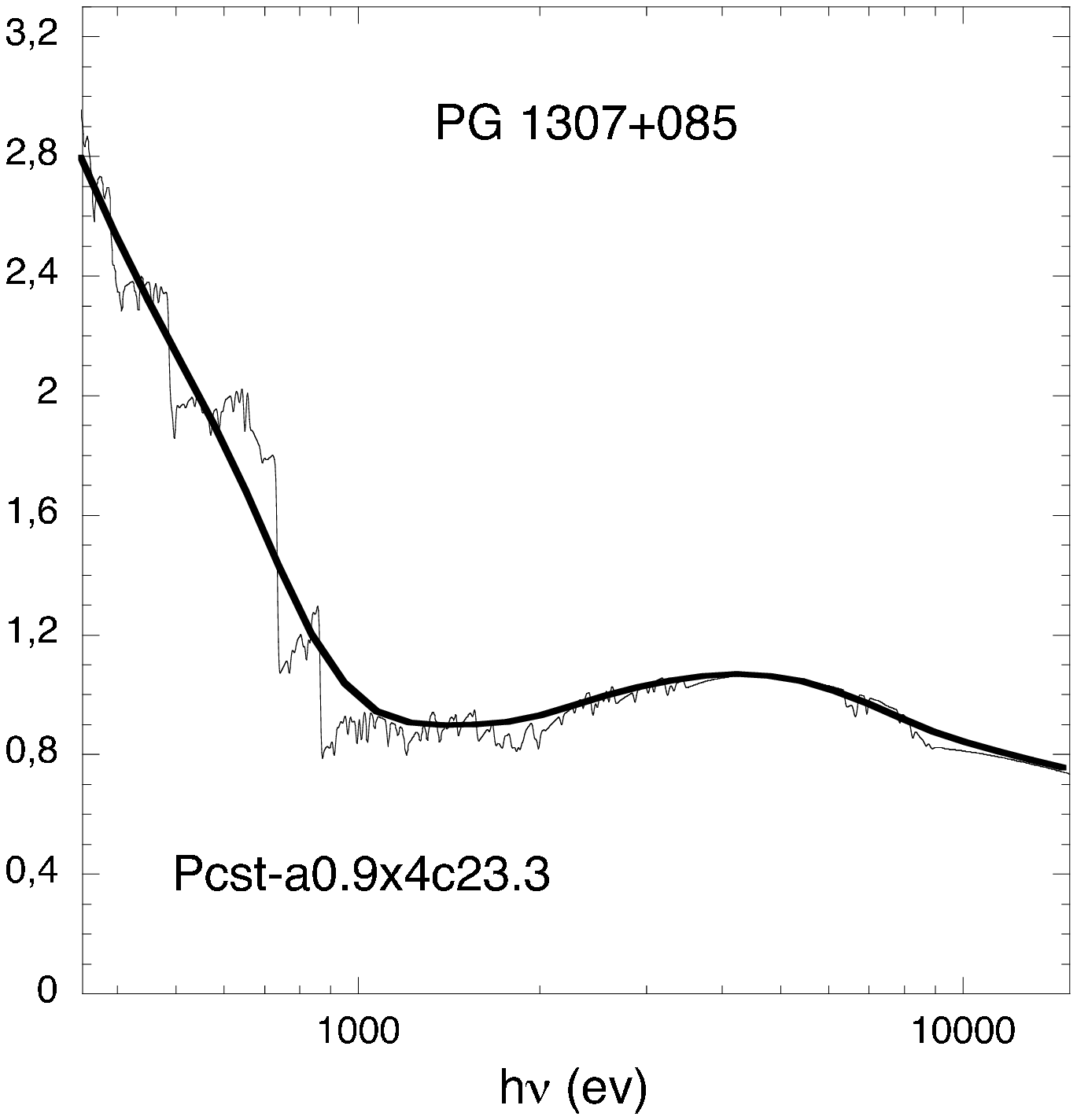}
\caption{Comparison between the observed and computed X-ray spectra for PG 1307+085, from the sample of Piconcelli et al. (2005, see their Fig. 2). The model is a constant total pressure slab  illuminated by a power law continuum with $\alpha=0.9$, $\xi=10^4$  and $N$= 2 10$^{23}$ cm$^{-2}$.
Top panel: the computed spectrum is displayed with a spectral resolution of 2. The observed and computed spectra have been both divided by a power law of photon index $\Gamma=1.5$.
Bottom panel: the computed absorption spectrum of the top panel is displayed with a spectral resolution of 2 (thick solid line), and 100 (thin solid line). The spectra have been divided by the same power law. It is extended up to 15 keV.}
 \label{fig-comp-mod-obs}
\end{figure}

As we can see, it is thus difficult to obtain simultaneously a strong X-ray excess and a relatively flat 2-10~keV slope - i.e. not  increasing too steeply. Moreover, a  ``wiggle" appears always in the 1-12 keV spectrum, due to the decrease of absorbing species above 2 keV and to the presence of iron edges above 7 keV. Therefore, a pure power law cannot be obtained in spite of the considerable smearing. With a dispersion velocity smaller than $v/c=0.2$, several other ``wiggles" are apparent. To illustrate the effect of the large dispersion velocity, the spectrum is also displayed on Fig. \ref {fig-comp-mod-obs} with a spectral resolution of 100 (FWHM $\sim 1300$ km.s$^{-1}$), corresponding typically to the velocities in the Broad Line Region and of a classical WA. It allows to see the numerous features which are smeared by the assumed gas motions.\\
Those ``wiggles" become apparent with a resolution bigger than 5 (i.e. a dispersion velocity less than $0.08 c$)}. It is clear that {\it a smaller velocity dispersion would be unable to account for the smoothness of the observed spectra.}

\begin{figure*}[tbp]
 \centering
 \includegraphics[width=15cm]{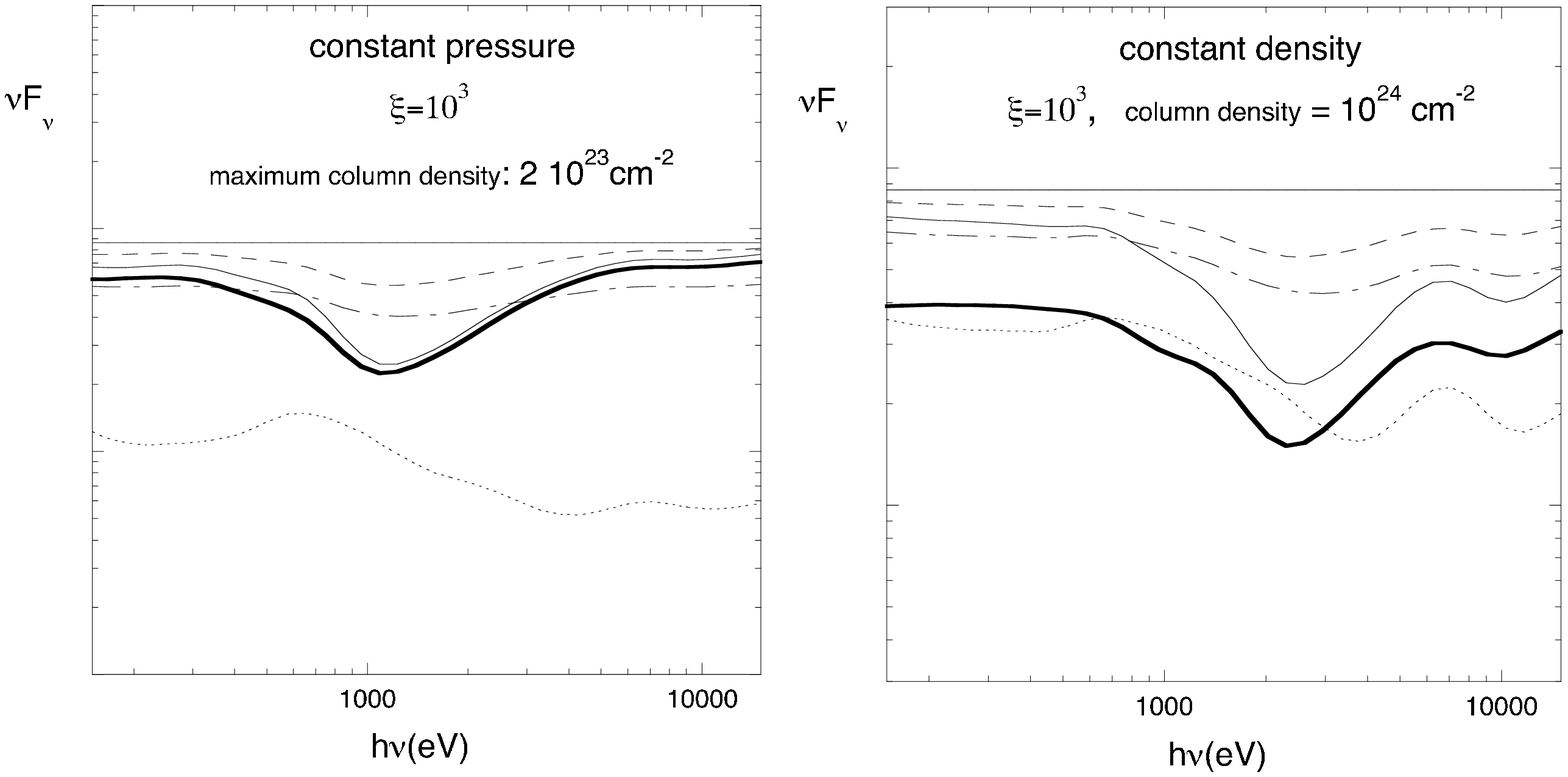}
\caption{ Different combinations of the absorbed, emitted, and reflected spectra are displayed with a spectral resolution of 2, for a constant total pressure model which has the maximum possible thickness for its given ionization parameter, 2 10$^{23}$ cm$^{-2}$, and for a constant density model with a larger column density and the same ionization parameter. Thick solid line: pure absorption spectrum. Thin solid lines: absorption + outward emission spectra, assuming that the absorbing medium is a spherical envelope around the primary source. Dotted lines: reflection spectrum. Dashed lines: combination of equal proportions of primary and absorption spectra (i.e. partial coverage of the line of sight equal to 50$\%$ ). Large and small dashed lines: combination of equal proportions of primary, reflection, and absorption spectra. The straight lines represent the primary continuum. Fluxes are in arbitrary units.}
 \label{fig-combinaisons-spe}
\end{figure*}
 
Note that a better fit could be obtained by summing up the contribution of two or more absorbing media, since one would then add two degrees of freedom for each new model ($\xi$ and the column density). The impact on the absorption spectrum depends also on the disposition of the absorbers: if they cover completely the primary source and are located at different distances from the source, the external absorber will be illuminated by an already absorbed spectrum; while if they are mixed together, they will receive the same incident spectrum. Therefore the degree of freedom would  be increased, but the constant total pressure requirement would still limit the strength of the X-ray absorption.

  \bigskip

Can we conclude that the pure absorption hypothesis can account for the soft X-ray excess in AGN?
We tried to fit other objects from Piconcelli et al. (2005) sample, without an agreement as good as with PG 1307+085.  There are at least two reasons for that: 

\begin{itemize}
\item  our grid of models is restricted, and does not take into account all possible combinations of parameters;
\item two observational parameters are insufficient to determine the shape of the whole X-ray spectrum.
\end{itemize} 

  Our purpose in a further work is thus to extend the grid of models and to define several other observational parameters, in order to get (through $\chi^2$ tests) a unique and good solution for each object.  Presently, though we can already state that - at the present uncertainty level of the observations, and in the range of energy from 0.3 to 10 keV - some spectra can be accounted for by pure absorption constant total pressure  models, it is not clear at all if all spectra could be accommodated by such models.

\section {Emission, partial covering, and reflection}

\subsection{Low column density: emission and partial covering}
\label{sect:mixed}

Until now, we have considered pure absorption spectra, assuming that the emission produced by the medium surrounding the primary source is negligible. This is true only if its covering factor and/or its column density are small. A small covering factor would be in contradiction with the high occurence of soft X-ray excess, assumed here to be caused by the absorbing medium.  It would thus be extremely unlikely that a large proportion of quasars and Seyfert nuclei display such an absorption, and that the absorbing medium would be confined to the line of sight of the primary source. Moreover, it is expected that the ``emission" coverage factor is about the same as that of the ``absorption" one, unless the medium has a very peculiar geometry.  So the absorbing medium should surround the source with a coverage factor of the order of unity. In this case, it does not only absorb, but also emit in the UV and soft X-ray range, according to the temperature reached by the deep layers close to the back (non-illuminated) surface. 	

This emission can be important for relatively thick media illuminated by an intense radiation field. Even for a relatively thin model, the ``total" spectrum including the emission (computed in assuming that the absorbing medium is spherically distributed around the source with a coverage factor of unity)  is slightly more intense than the pure absorption one. They would be more different in the absence of the strong smearing, since several narrow spectral features would be present in emission in the total spectrum, while they are absent in the pure absorption spectrum. With a resolution of 2, the difference between the two spectra would not be detectable, since the primary continuum itself is not observed. 

\begin{figure*}[tp]
 \centering
 \includegraphics[width=15cm]{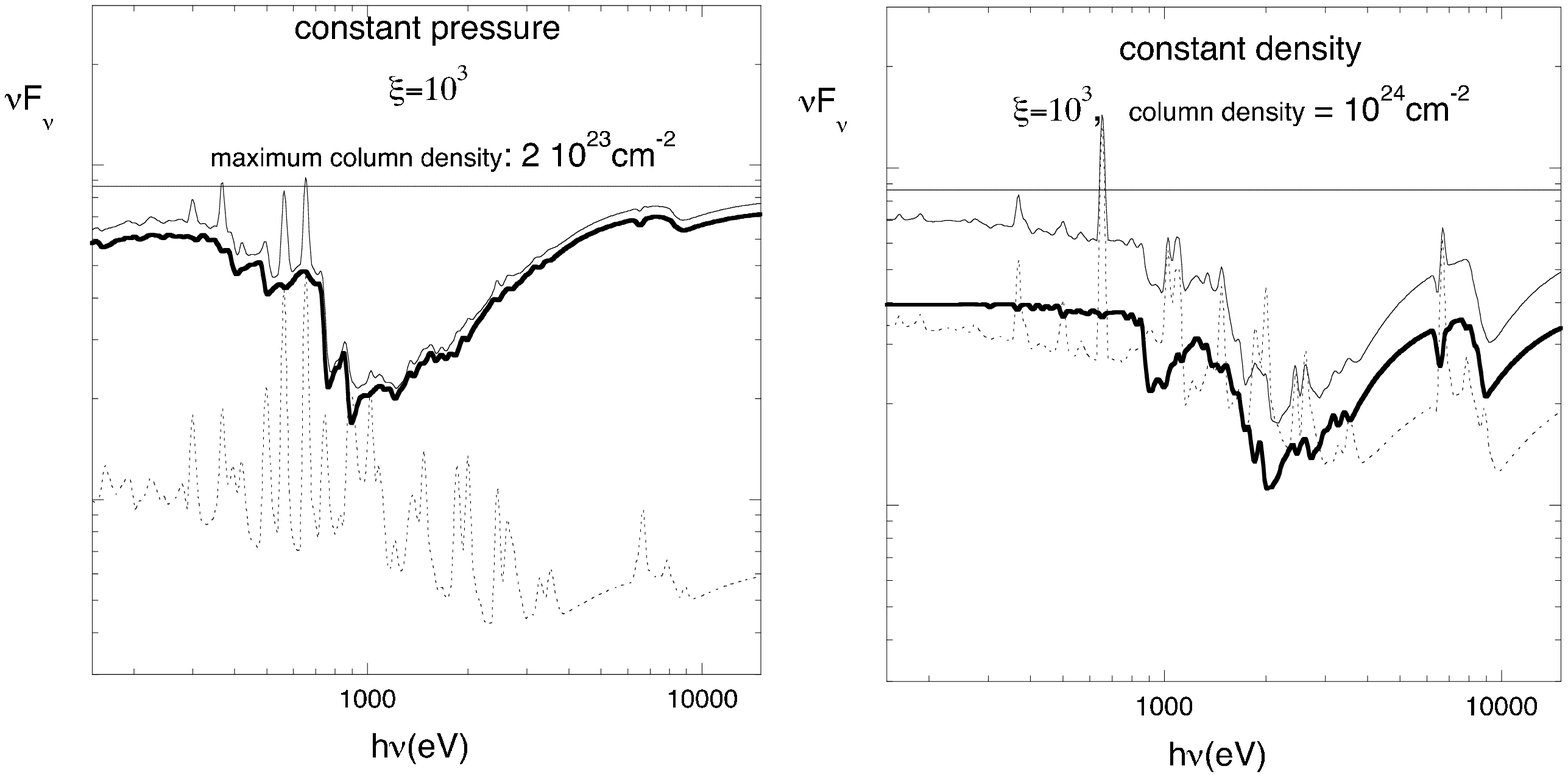}
\caption{ The spectra corresponding to the same models as on Fig. \ref {fig-combinaisons-spe} but displayed with a spectral resolution of 30. Thick solid lines: pure absorption spectrum. Thin solid lines: absorption + emission spectrum, assuming that the absorbing medium is a spherical envelope around the primary source. Dotted lines: reflection spectrum.  The straight lines represent the primary continuum. Fluxes are in arbitrary units.}
 \label{fig-speR30}
\end{figure*}

Figure \ref{fig-combinaisons-spe} shows two examples of the influence of the emission, assuming that the absorbing medium is spherically distributed around the source with a coverage factor of unity. The left panel shows the spectrum of a constant total pressure model with its maximum thickness. Since this thickness is limited to a low value, the difference between the total and the pure absorption spectrum is small. The right panel shows the spectrum of a constant density medium with a large column density. Here the total spectrum differs appreciably from the pure absorption spectrum, and {\it the emission should be taken into account in the fitting procedure}.  Fig. \ref{fig-speR30} displays the same spectra but with a velocity dispersion equal to $c/75$ (corresponding to a spectral resolution of 30). Now the differences between the emission and absorption spectra appear distinctly, as the emission spectrum contains several intense emission lines which are not present in the pure absorption spectrum.	

Another effect not considered so far is a partial and not complete coverage factor of the absorbing medium. This possibility offers an enormous degree of freedom for the shape of the spectrum, and in particular it can erase efficiently the big absorption trough created in the pure absorption models. As an illustration Fig. \ref{fig-combinaisons-spe} shows (in dashed lines) the result of a partial coverage of 50$\%$ and one can see that the absorption is almost suppressed. It is clear that keeping the coverage factor as a free parameter would allow to fit all kinds of spectra, when added to the other important parameters, i.e. the spectral index of the primary continuum, the ionization parameter and the column density. 

Finally, if the primary source is not completely obscured by the absorbing medium we have also to take into account another ingredient in the observed spectrum, namely the reflection spectrum. Since the whole covering factor is smaller than unity, one can possibly see also the side of the absorbing clouds facing the primary source. Of course, it implies that the primary source itself is not opaque, or that it is intimately mixed with the absorbing/emitting medium. 

\subsection {High column density: reflection models} 

\begin{figure}
 \centering
 \includegraphics[width=9cm]{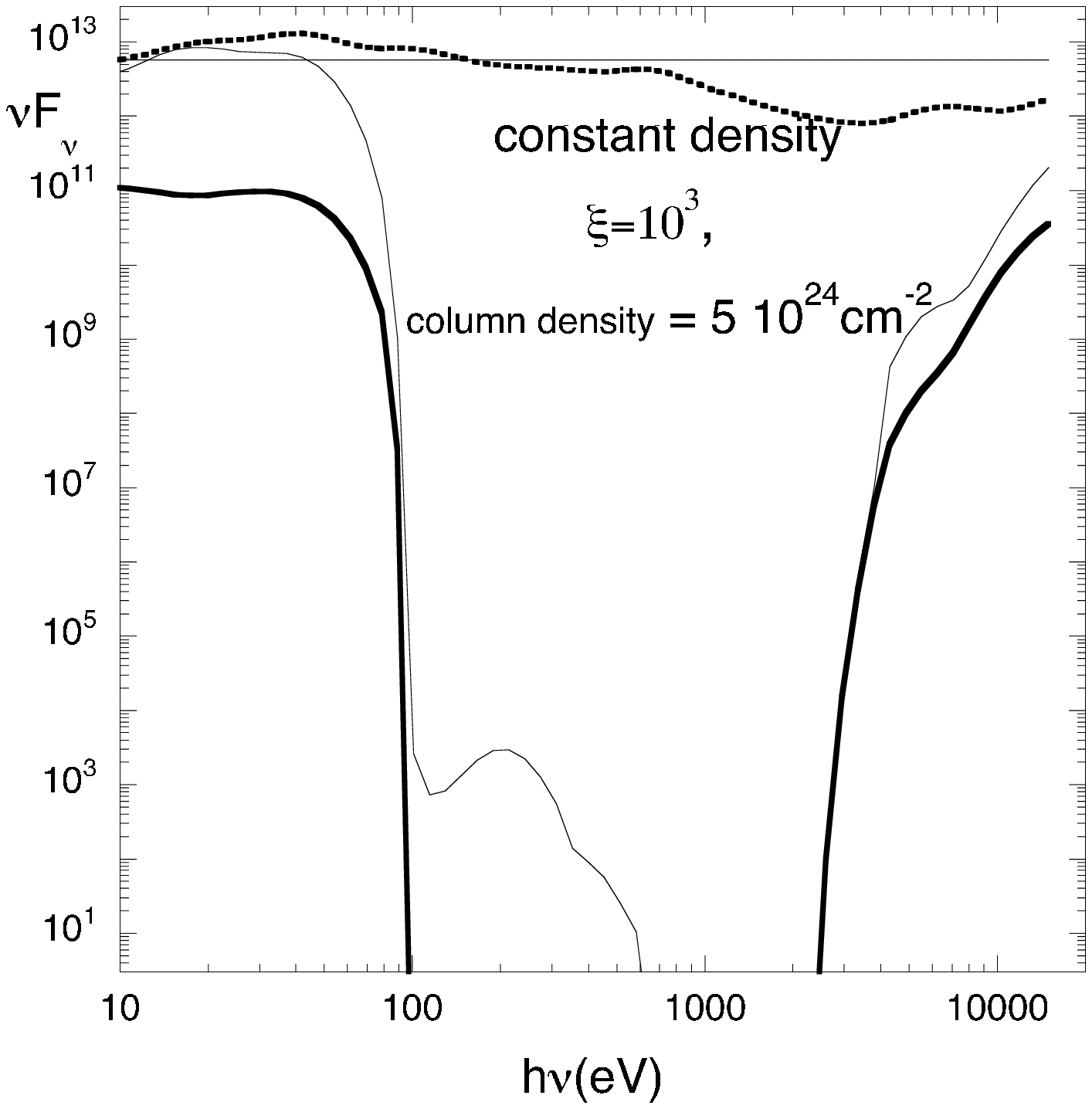}
\caption{ Spectrum for a constant density thick model, and the standard primary continuum  displayed with a spectral resolution of 2. Thick solid lines: pure absorption spectrum. Thin solid lines: absorption + emission spectrum, assuming that the absorbing medium is a spherical envelope around the primary source. Thick dotted lines: reflection spectrum.  The straight line represents the primary continuum. Fluxes are in arbitrary units.}
 \label{fig-epais-ref-abs-em}
\end{figure}

Up to now we have considered relatively thin media, whose Thomson thickness is smaller than unity.
Let us consider now thick media with a Thomson thickness of the order of or larger than  unity (i.e., a column density larger than 10$^{24}$ cm$^{-2}$).  If a thick medium covers completely the line of sight of the primary source, nothing except the extreme parts of this continuum (the UV and the Gamma bands) would be seen, as shown on Fig. \ref{fig-epais-ref-abs-em}. We enter here in the domain of the ``reflection" models, which were invoked to account for the presence of the soft X-ray band. 

\begin{figure}
 \centering
 \includegraphics[width=9cm]{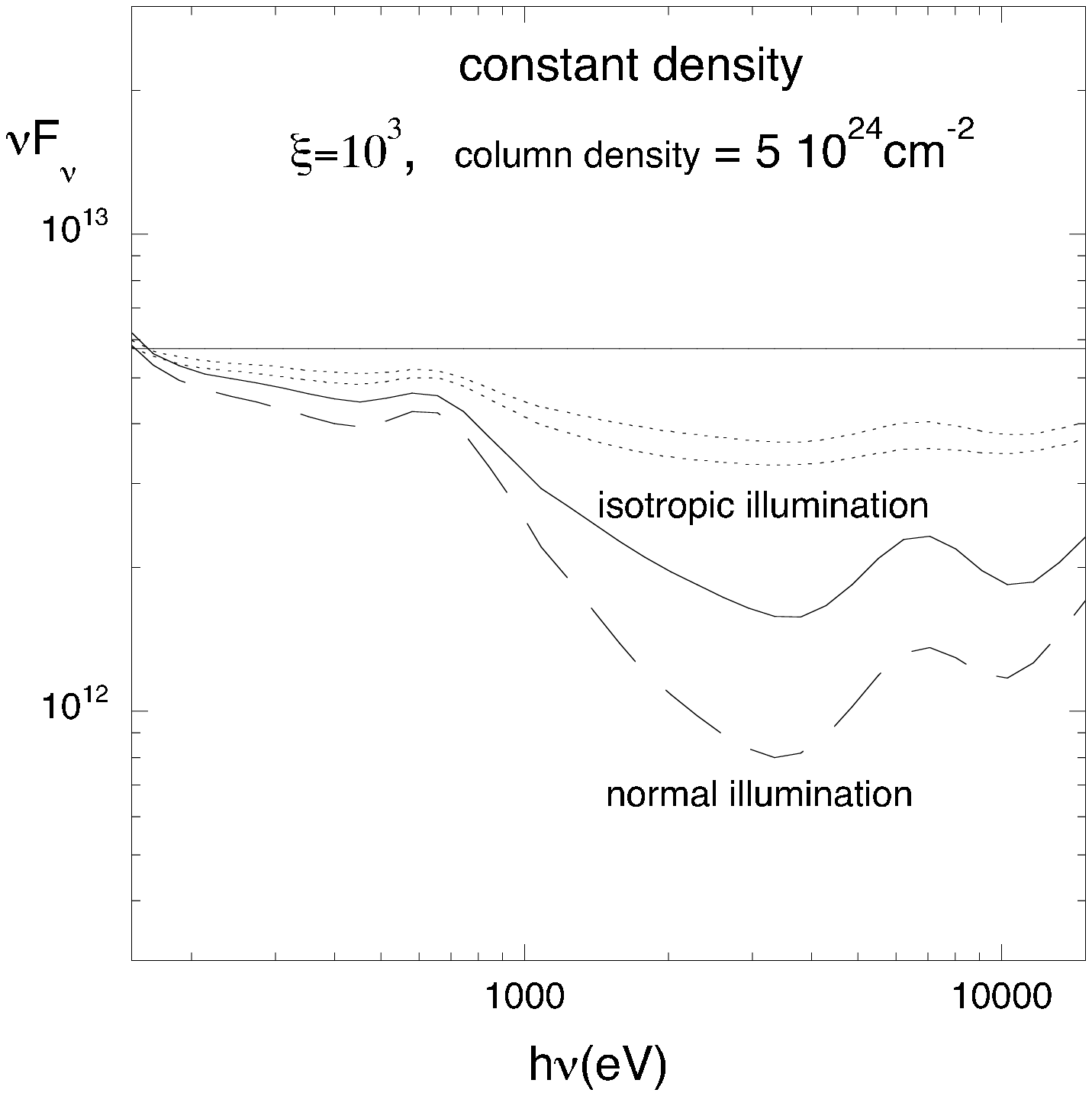}
\caption{ A zoom on the reflection spectrum for the thick model given on Fig. \ref{fig-epais-ref-abs-em}. The solid and long dashed lines show the reflection spectra in the case of a normal and an isotropic illumination, and the dotted lines shows the ``observed spectrum", i.e. the half sum of the reflection and primary spectrum.  The straight line represents the primary continuum. Fluxes are in arbitrary units.}
 \label{fig-comp-iso-norm-spe}
\end{figure}

\begin{figure}[tb]
 \centering
 \includegraphics[width=9cm]{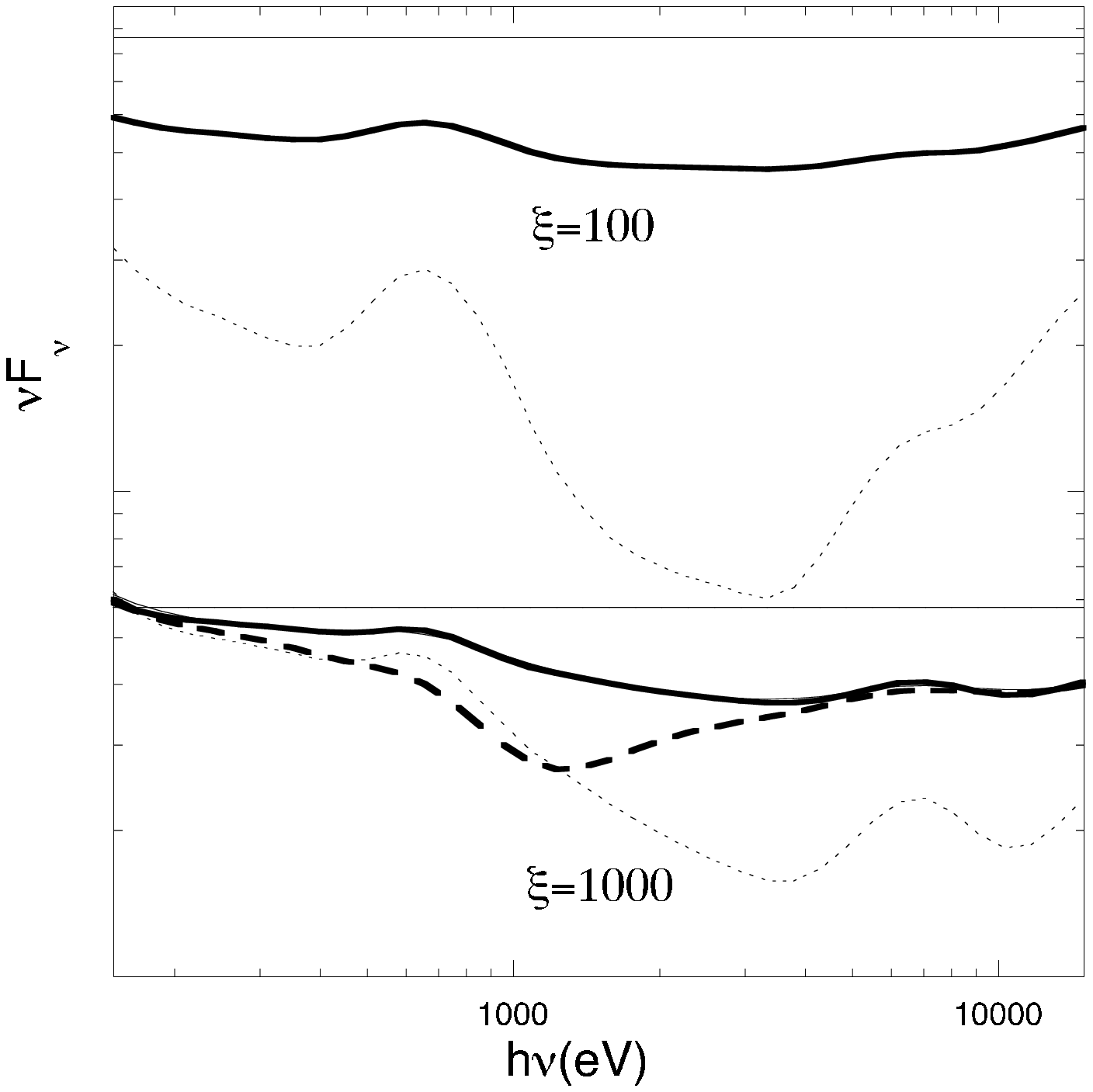}
\caption{ Computed spectra for thick reflection models for two values of the ionization parameter. The straight lines represent the primary continuum. Thin dotted lines: reflection spectrum. Thick solid lines: half sum of reflection plus primary spectrum. The figure illustrates the fact that the spectrum which reaches the observer (reflection plus primary) never displays a strong X-ray excess. The thick dashed line shows one of the ``observed" spectrum, after being absorbed by a constant total pressure slab of column density 10$^{22}$ cm$^{-2}$ and an ionization parameter $\xi=100$, with a dispersion velocity equal to 0.2$c$. Fluxes are in arbitrary units.}
 \label{fig-x2x3-spe}
\end{figure}

 In almost all the cases considered so far, the reflection spectrum was negligible, as it can be seen in the left panel on Fig. \ref{fig-combinaisons-spe}, where the reflection spectra are shown, as well as a combination of equal proportions of primary, absorption, and reflection spectra. For thick ionized media, reflection dominates completely the transmission and the outward  emission. But it is not seen, unless the surrounding medium is located {\it behind or sideways} the primary source. The atmosphere of an accretion disk illuminated from above, either by a central source or by a patchy hot corona (Capriotti et al. 1981, Haardt \& Maraschi 1991, 1993, Ross \& Fabian 1993, Nayakshin et al. 2000, Nayakshin \& Kallman 2001), is considered as the best location for such a reflection.  The disk atmosphere is then in hydrostatic equilibrium under the gravity of the central black hole, but it was often represented by a constant density slab (Ross \& Fabian 1993 and subsequent works), or by a slab in total pressure equilibrium (Collin et al. 2003), in both cases  heated from below by the viscous release of gravitational energy. 
 
We have run a set of constant density thick models to check whether it is possible to account for the soft X-ray excess in PG quasars by reflection. Instead of the normal illumination considered previously, we assumed an isotropic illumination, which is more realistic if the reflecting medium is located close to the primary source and has a comparable size (it was actually the type of illumination considered in all previous studies performed with the TITAN code). As an illustration, Fig. \ref{fig-comp-iso-norm-spe} shows a comparison between the reflection spectra in the case of normal and isotropic illuminations, and we see that, as expected (because the mean path of the photons in the normal direction is smaller in the case of an isotropic illumination), the reflection is more important in the isotropic case. 	
 
 Figure \ref{fig-x2x3-spe} shows two examples of such models, with a small and with a large ionization parameter. 
 It is obvious that when the reflection spectrum is added to the primary continuum, the soft X-ray excess almost disappears, and one understands why it is so difficult, with such models, to account for the large soft excess observed in some PG quasars and in high accretors like NLS1s.
Actually we have tried without any success to fit some other observed spectra from the Piconcelli et al. (2005) sample, when the primary continuum was added to the reflection one. To summarize the problem:
 
  \begin{itemize}
 \item either the ionization parameter is small and the reflection spectrum displays a strong X-ray excess, but it is negligible compared to the primary one; 
 \item or the ionization parameter is large, and the reflection spectrum is comparable in flux to the primary one, but it has only a very small X-ray excess.
 \end{itemize}
 
 It is why a new disk model has been proposed by Fabian et al. (2002). They consider a disk made of a dilute hot gas creating a hard X-ray spectrum which is not seen directly, and whose radiation is reflected on sheets of dense material formed by disk instabilities. Thus the primary source is hidden and only the reflection spectrum is observed. This is presently the only available reflection model able to explain large X-ray excesses, but unfortunately it is valid only  for objects accreting close to the Eddington rate, like NLS1s.
 An alternative approach to hide the primary continuum could be due to the effect of ``light bending'' where a significant fraction of light emitted by the central source is bent onto the disc rather than escaping to the observer (Crummy et al. 2005).

\section{Discussion}		

 In summary, absorption models seem able to account for some of the soft X-ray excesses, and we do not know whether one could find good fits for the others with a better grid of models. A mixture of absorbing clouds spanning a range of optical thicknesses and ionization parameters, or covering only partially the primary source, would also help to accommodate all observed spectra. On the contrary, reflection models underpredict even relatively modest soft X-ray excesses, unless the primary continuum is hidden. 
 
It is thus important to examine what are the implications of these models. Both absorption and reflection models require large ionization parameters and large values of the column density. 
But {\it the most important condition is the smearing of the spectral features by a very large dispersion velocity}. We have indeed seen that a velocity dispersion smaller than $c/5$  cannot account for the smoothness of the observed spectrum. This large velocity implies that :

\begin{itemize}
\item either the absorbing/reflecting medium is located close to the black hole and is thus mixed with the primary source; this would be the case of a medium dominated by rotation and turbulent motions,
\item or it is located further out, but in this case it is not  gravitationally bound to the black hole and it is necessarily in outflowing motion.
\end{itemize}

 A likely solution for the overall X-ray emission of AGN is thus a hot medium emitting a primary hard X-ray continuum, surrounded by or embedded into a system of absorbing/reflecting/emitting photoionized clouds. In this hypothesis, the dominance of the absorption, or the emission, or the reflection spectrum, would be determined by the coverage factor of the primary source by the clouds, and by their thickness. This clumpy medium could be either a wind dominated by outflowing motions, or a thick inhomogeneous accretion flow.
 
\medskip

 Let us first assume that the high velocity is due to an outflowing motion. Note that in this case the constant total pressure models considered so far would no more be valid, as the medium would be in a dynamical state. So possibly our constraint on the thickness due to the thermal instability would not hold, but it would be replaced by other constraints, such as those due to radiation pressure acceleration, which would also limit the column density of the absorbing matter.  Second, the ``dispersion velocity" (actually a velocity gradient) would be accompanied by a blueshift of the lines, but it would not change appreciably the shape of the absorption spectrum. 
 
 	The wind model is actually reminiscent of the recent discovery of several quasars showing extreme absorption properties (in particular blue-shifted highly ionized Fe lines and a sharp feature at 7 keV which is attributed to the K-shell edge), with outflowing velocities of a few tenths of the light speed, and column densities of the order of  $10^{24}\ \mathrm{cm}^{-2}$ (Pounds et al. 2003a and 2003b, 2004). The ionization parameter at the surface of this medium is of the order of $10^3$, like in our models. The question is raised of whether these absorbers could be a general feature of AGN which is still undetected, as the highly ionised gas is hard to observe. 
 
Assuming the stationary outflow to be a relatively thin shell as compared to its distance from the central black hole, one can estimate the mass outflowing rate $\dot{M}_{\rm out}$:  

\begin {equation}
\dot{M}_{\rm out}\sim \Omega \ f_{\rm vol} \ m_{\rm H}\ V_{\rm out}\ n_\mathrm{H}\ R^2,
\label{eq-ouflow1}
\end{equation}
where $\Omega$ is the opening angle of the outflowing medium, $f_{\rm vol}$ is the volume filling factor (if the medium is made of small clumps), $V_{\rm out}$ the outflowing velocity,  $m_{\rm H}$ the hydrogen mass. Dividing the outflowing rate  by the Eddington accretion rate $\dot{M}_{\rm Edd}$, and using our definition of $\xi=L/n_\mathrm{H}R^2$, one gets:

\begin {equation}
{\dot{M}_{\rm out}\over \dot{M}_{\rm Edd}}\sim { \Omega f_{\rm vol} \  m_{\rm H}\ V_{\rm out}\ R_{\rm Edd}\ \eta\ c^2 \over \xi}
\label{eq-ouflow2}
\end{equation}
where $R_{\rm Edd}$ is the Eddington ratio $L/L_{\rm Edd}$ and $\eta$ is the mass-energy efficiency conversion factor. If we assume that the whole dispersion velocity is due to outflowing motions with an opening angle of the order of $0.5 \times 4\pi \sim 6.3$ if one admits that the number of Seyfert 1 = 50\% of Seyfert 2, we get:

\begin {equation}
{\dot{M}_{\rm out}\over \dot{M}_{\rm Edd}}\sim 2\ 10^3 \ f_{\rm vol}\   \left[{R_{\rm Edd}\over 0.3}\right]  \left[{V_{\rm out}\over 0.2c}\right]\ \left[{\eta \over 0.1}\right]\ \left[{10^3\over \xi}\right].
\label{eq-ouflow3}
\end{equation}

Eq. \ref{eq-ouflow3} shows that $f_{\rm vol}$ should be smaller than $\sim 5\ 10^{-4}$, in order for the outflowing rate not to have an unrealistic value much larger than the Eddington rate. 

It is interesting to deduce other physical properties of the absorbing medium. Since $n_\mathrm{H}\ f_{\rm vol}\ \Delta R$ is roughly equal to the column density $N$, where $\Delta R$ is the geometrical thickness of this outflow whose base is at distance $R$ from the central regions, and using the constraint $\Delta R / R < 1$, we obtain for the distance $R$:

\begin {equation}
R < {f_{\rm vol}\   R_{\rm Edd}\ L_{\rm Edd} \over \xi \ N},
\label{eq-ouflow4}
\end{equation}
or:
\begin {equation}
{R\over R_{\rm G}} <  10^6 \ f_{\rm vol}\   \left[{R_{\rm Edd}\over 0.3}\right] \left[{10^3\over \xi}\right] \left[{3\ 10^{23}\over N}\right].
\label{eq-ouflow5}
\end{equation}

As $f_{\rm vol}$ should be smaller than $\sim 5\ 10^{-4}$,  $R/R_{\rm G} < 500$.
From our definition of $\xi$, we can also find the value of the density:

\begin {equation}
n_\mathrm{H} > 8\ 10^{11} \left[{R_{\rm Edd}\over 0.3}\right]
\left [ \frac{10^{3}}{\xi} \right ]
\left [ \frac{10^7 M_{\sun}}{M} \right ]
\left [ \frac{500 R_{\rm G}}{R} \right ] ^2\ \ {\rm cm}^{-3}.
\label{eq-ouflow10}
\end{equation}
Thus the density should be high, actually of the order of that of the atmosphere of a standard accretion disk. 

On the other hand, the spread of velocity required by the absorption model implies the existence of many clouds on the line of sight, with a smooth velocity gradient up to $0.2c$ (FWHM = 60~000 km\ s$^{-1}$). The thermal velocity of hydrogen nuclei being of the order of 100 km\ s$^{-1}$ in the gas giving rise to the absorption spectrum, it means that at least 600 clouds on a line of sight  are required to give a dispersion velocity equal to 0.2$c$. The ratio $ f_{\rm vol}/ f_{\rm cov}$ is of the order of $r_{\rm c} N_{\rm los}/\Delta R= N/(n_\mathrm{H}\Delta R)$, where $r_{\rm c}$ is the dimension of a cloud, and $N_{\rm los}$ the mean number of clouds on a line of sight. For $N_{\rm los}$= 600, one gets $r_{\rm c} < 6\ 10^8 \left[N\over {3\ 10^{23}}\right] \left[{8\ 10^{11}\over n_\mathrm{H}}\right]$ cm. Finally, since $f_{\rm cov} \sim 1$ (primary source completely covered), $ f_{\rm vol}/ f_{\rm cov}$ should be smaller than $5\ 10^{-4}$ and the following condition should be fullfilled:

\begin {equation}
\left [ \frac{8\ 10^{11}}{n_\mathrm{H}} \right ] \left[N\over {3\ 10^{23}}\right] \left [ \frac{10^7 M_{\sun}}{M} \right ] \left [ \frac{500 R_{\rm G}} {R}\right ]  > 1.
\label{eq-ouflow11}
\end{equation}

We see that the inequality (\ref{eq-ouflow11}) is marginally realized.
Thus the model is consistent from the phenomenological point of view. It stays however to be justified on a physical ground, especially to seek for a confining mechanism of the small dense clouds (magnetic?). Note that they could be simply transient entities.

As the absorbing medium should be located inside 500 $R_{\rm G}$, its dynamics should comprise a fraction of rotational motion. It could perhaps be identified with the inner part of  a radiatively driven wind  launched by  the disk, as proposed by Murray \& Chiang (1995, and subsequent works). However the wind does not reach
relativistic velocities in their model. The high velocities close to the black hole could also be triggered by the accretion disk releasing magnetically driven winds, like those proposed e.g. by K\"onigl \& Kartje (1994).  In particular Czerny \& Goosmann (2004) showed that external X-ray heating by magnetic flares can account for vertical acceleration of disk material.

In this computation we have neglected the medium at a larger distance. The stationary assumption indeed implies the presence of matter at any distance from the center. Since the column density of this medium decreases as $R^{-2}$, the material located at, say, 1000 $R_{\rm G}$, would have a column density 100 times smaller than this value inside 100 $R_{\rm G}$ and would therefore be undetectable (or detectable as a ``classical" Warm Absorber). 
Note also that we have assumed a stationary outflow, but of course it would have been equivalent to assume sporadic events, separated by the time (of the order of $R/V_{\rm out}$) that it would take for the absorbing medium to be replenished before its disappearence by dilution. 

\medskip

An alternative solution to the outflow is a thick inhomogeneous accretion flow, whose dynamics is dominated by rotation and turbulent motions.  But now  the absorbing medium should be located at a distance of the order of 25~$R_{\rm G}$, as it should be gravitationally bound to the black hole and simultaneously have relativistic velocities. 
This is actually  close to the model proposed by Collin et al. (1996), consisting in a quasi-spherical (or a thick disk) distribution of clouds covering almost totally a primary source of X-rays. The difference with the present model lies in the very large column density of the Collin et al. accretion flow (Thomson opacity of the order of 10-100), which did not allows the leakage of any transmitted radiation, so partial covering was required and had to be fine tuned. The interest of this model was to account not only for the X-ray emission, but also for the UV spectrum, due to the outward emission of the optically thick cloud system. In the present model, another UV source is required, like a geometrically thin accretion disk.

\medskip
 
Finally, none of the models are very satisfactory from a physical point of view. Owing to its large column density, the wind implies too massive outflows. Both the wind and the accretion models require an additional UV emission, which has to be provided by a geometrically thin accretion disk. The coexistence of a spherical accretion flow and a thin disk seems quite artificial.

 So we would prefer an ``hybrid model", including an accretion disk with a hot patchy corona emitting the whole X-ray spectrum, through the ``classical" Compton reflection on the disk and inverse Compton process in the corona initially proposed by Haardt \& Maraschi (1991 and 1993), the UV emission being produced by the viscous release in the disk. The emerging spectrum in the 0.1 - 10 keV range would be similar to the ``observed" spectra shown on Fig. \ref{fig-x2x3-spe} and would display a small soft X-ray excess. If such a spectrum is absorbed by a moderately thick wind, the excess would be increased and would become comparable to the observations. As an illustration, Fig. \ref{fig-x2x3-spe} shows one of the ``observed" spectra, after being absorbed by a constant total pressure slab of column density 10$^{22}$ cm$^{-2}$ with an ionization parameter $\xi=100$. Assuming that this slab has also a dispersion velocity equal to 0.2$c$, one gets a spectrum quite comparable to the observations. The big interest of this hybrid model is to require less extreme conditions for the outflowing mass, as the column density and the ionization parameter of the wind are smaller. According to Eqs. \ref{eq-ouflow3} and \ref{eq-ouflow5}, $f_{\rm vol}$, and thus $\dot{M}_{\rm out}/ \dot{M}_{\rm Edd}$, would be 300 and 30 times smaller than in the previous absorption model, respectively.

\section{Conclusion}

This study has shown that absorption models could account for some strong soft X-ray excesses, while reflection models are possibly able to account for weak soft X-ray excesses, but certainly not for the large ones (including those of typical PG quasars), unless the primary continuum is hidden from our view.  An important conclusion is also that pure absorption models require a kind of ``fine tuning" of the absorber, in order to constraint the 1 keV trough, which otherwise could have any strength. We have suggested a medium in total pressure equilibrium, which leads to a maximum intensity of the trough, as well as a ``universal" shape of this maximum trough, due to the thermal instability mechanism. A complete grid of constant total pressure models, very demanding in computation time, is necessary to pursue this study.  

In the absorption model, either a thick accretion flow, or a relativistic wind is required. None of them seem very realistic from a physical point of view, and moreover  both models require an additional source of UV emission, like a geometrically thin accretion disk. Therefore we favor an ``hybrid" model, where the primary UV-X source could be produced by a disk-corona system, and then absorbed by a modest relativistic wind.

A clue to the problem could be obtained with observational data from Astro-E2, as this instrument should give good spectra above 10 keV. Looking at Fig. \ref{fig-comp-mod-obs}, where plotting the energy up to 15 keV reveals a well--defined slope, this might bring important limitations. 
Variability is also a key point. It can help to disantangle the absorption and reflection models, as one would expect different behaviours of the light curves in the hard and soft X-ray bands: in the first case, the hard and the soft X-ray flux should vary simultaneously, while in the reflection model, there should be a time delay between the two light curves. 

\begin{acknowledgements}
A.\,C. Gon\c{c}alves acknowledges support from the {\it {Funda\c{c}\~ao para a Ci\^encia e a Tecnologia}}, Portugal, under grant no. BPD/11641/2002. Part of this work was supported by the Laboratoire Europ\'een Associ\' e Astrophysique Pologne-France.
\end{acknowledgements}


\begin{thebibliography}{99}

\bibitem[Barvainis(1993)]{1993ApJ...412..513B}
Barvainis, R.\ 1993, \apj, 412, 513 

\bibitem[Behar et al.(2003)]{2003ApJ...598..232B}
Behar, E., Rasmussen, A.~P., Blustin, A.~J., Sako, M., Kahn, S.~M., Kaastra, J.~S., Branduardi-Raymont, G., \& Steenbrugge, K.~C.\ 2003, \apj, 598, 232 

\bibitem[Blustin et al.(2005)]{2005A&A...431..111B}
Blustin, A.~J., Page, M.~J., Fuerst, S.~V., Branduardi-Raymont, G., \& Ashton, C.~E.\ 2005, \aap, 431, 111 

\bibitem{branduardi2001} 
Branduardi-Raymont, G., Sako, M., Kahn, S.~M., Brinkman, A.~C., Kaastra, J.~S., \& Page, M.~J.\ 2001, A\&A, 365, L140 

\bibitem[Capriotti et al. 1981, ApJ, 245, 396 ]{capriotti1981}
Capriotti, E., Foltz, C., \& Byard, P.\ 1981, ApJ, 245, 396

\bibitem[Collin-Souffrin et al.(1996)]{1996A&A...314..393C} 
Collin-Souffrin, S., Czerny, B., Dumont, A.-M., \& Zycki, P.~T.\ 1996, \aap, 314, 393  

\bibitem[Collin et al.(2003)]{2003A&A...400..437C}
Collin, S., Coup{\' e}, S., Dumont, A.-M., Petrucci, P.-O., \& R{\' o}{\. z}a{\' n}ska, A.\ 2003, \aap, 400, 437 

\bibitem[Collin et al.(2004)]{2004A&A...419..877C}
Collin, S., Dumont, A.-M., \& Godet, O.\ 2004, \aap, 419, 877 

\bibitem{crummy2005}
Crummy, J., Fabian, A. C., Brandt, W. N., \& Boller, Th.\ 2005, \mnras, 361, 1197

\bibitem{czerny1994}
Czerny B., \&. Zycki, P.T. 1994, ApJ, 431, L5

\bibitem{2004A&A...428...353C}
Czerny, B., \& Goosmann, R.\ 2004, \aap, 428, 353
  
\bibitem[Dumont et al.(2000)]{2000A&A...357..823D}
Dumont, A.-M., Abrassart, A., \& Collin, S.\ 2000, \aap, 357, 823 

\bibitem[Dumont et al.(2003)]{2003A&A...407...13D}
Dumont, A.-M., Collin, S., Paletou, F., Coup{\' e}, S., Godet, O., \& Pelat, D.\ 2003, \aap, 407, 13 

\bibitem[Fabian et al.(2002)]{2002MNRAS.331L..35F}
Fabian, A.~C., Ballantyne, D.~R., Merloni, A., Vaughan, S., Iwasawa, K., \& Boller, T.\ 2002, \mnras, 331, L35 

\bibitem{ferlan1998}
Ferland, G.J., Korista, T., Verner, D.A., Ferguson, J.W., Kingdom, J.B., \& Verner, E.M. 1998, PASP, 110, 761

\bibitem{gallagher2002}
Gallagher S.C., Brandt W.N., Chartas G., \& Garmire G.P., 2002, ApJ, 567, 37

\bibitem{gallo2004}
Gallo L.C., Tanaka Y., Boller Th., Fabian A.C., Vaughan S., \& Brabdt W.N. 2004, MNRAS, 353, 1064
 
 \bibitem[Gierli{\' n}ski \& Done(2004)]{2004MNRAS.347..885G}
 Gierli{\' n}ski, M., \& Done, C.\ 2004, \mnras, 349, L7 

\bibitem{grupe2003}
Grupe D., Mathur S., \& Elvis M. 2003, AJ, 126, 1159

\bibitem[Halpern(1984)]{1984ApJ...281...90H}
Halpern, J.~P.\ 1984, \apj, 281, 90 

\bibitem[Haardt \& Maraschi(1991)]{1991ApJ...380L..51H}
Haardt, F., \& Maraschi, L.\ 1991, \apjl, 380, L51 

\bibitem[Haardt \& Maraschi(1993)]{1993ApJ...413..507H}
Haardt, F., \& Maraschi, L.\ 1993, \apj, 413, 507 

\bibitem[Ives et al.(1976)]{1976ApJ...207L.159I} Ives, J.~C., Sanford, 
P.~W., \& Penston, M.~V.\ 1976, \apjl, 207, L159 

\bibitem[Kaastra et al.(2002)]{2002A&A...386..427K}
Kaastra, J.~S., Steenbrugge, K.~C., Raassen, A.~J.~J., van der Meer, R.~L.~J., Brinkman, A.~C., Liedahl, D.~A., Behar, E., \& de Rosa, A.\ 2002, \aap, 386, 427 

\bibitem{kallman1995}
Kallman, T.R., \& Krolik, J.H. 1995, XSTAR, a Spectral Analysis Tool, Users Guide

\bibitem{kinkhabwala2002}
Kinkhabwala A., Sako M., Behar E. Kahn S.M., Paerels F., \& Brinkman, A.C. 2002, ApJ, 575, 732 

\bibitem{1994ApJ...434...446K}
K\"onigl, A., \& Kartje, J.~F.\ 1994, \apj, 434, 446

\bibitem[Krolik et al.(1981)]{1981ApJ...249..422K}
Krolik, J.~H., McKee, C.~F., \& Tarter, C.~B.\ 1981, \apj, 249, 422 
 
\bibitem[Laor et al.(1997)]{1997ApJ...477...93L}
Laor, A., Fiore, F., Elvis, M., Wilkes, B.~J., \& McDowell, J.~C.\ 1997, \apj, 477, 93 

\bibitem[Leighly et al.(1996)]{1996ApJ...469..147L}
Leighly, K.~M., Mushotzky, R.~F., Yaqoob, T., Kunieda, H., \& Edelson, R.\ 1996, \apj, 469, 147

\bibitem[Magdziarz et al.(1998)]{1998MNRAS.301..179M}
Magdziarz, P., Blaes, O.~M., Zdziarski, A.~A., Johnson, W.~N., \& Smith, D.~A.\ 1998, \mnras, 301, 179 

\bibitem[Murray \& Chiang(1995)]{1995ApJ...454L.105M}
Murray, N., \& Chiang, J.\ 1995, \apjl, 454, L105 

\bibitem[Mushotzky et al.(1978)]{1978ApJ...220..790M} Mushotzky, R.~F., 
Serlemitsos, P.~J., Boldt, E.~A., Holt, S.~S., \& Becker, R.~H.\ 1978, 
\apj, 220, 790 

\bibitem[Netzer(1993)]{1993ApJ...411..594N}
Netzer, H.\ 1993, \apj, 411, 594 

\bibitem[Netzer(1996)]{1996ApJ...473..781N}
Netzer, H.\ 1996, \apj, 473, 781

\bibitem[Nayakshin et al.(2000)]{2000ApJ...537..833N}
Nayakshin, S., Kazanas, D., \& Kallman, T.~R.\ 2000, \apj, 537, 833 

\bibitem[Nayakshin \& Kallman(2001)]{2001ApJ...546..406N}
Nayakshin, S., \& Kallman, T.~R.\ 2001, \apj, 546, 406

\bibitem[Nicastro et al.(2000)]{2000ApJ...536..718N}
Nicastro, F., et al.\ 2000, \apj, 536, 718 

\bibitem[P{\' e}quignot et al.(2001)]{2001ASPC..247..533P}
P{\' e}quignot, D., et al.\ 2001, ASP Conf.~Ser.~247: Spectroscopic Challenges of Photoionized Plasmas, 247, 533 

\bibitem[Piconcelli et al.(2004)]{2004MNRAS.351..161P}
Piconcelli, E., Jimenez-Bail{\' o}n, E., Guainazzi, M., Schartel, N., Rodr{\'{\i}}guez-Pascual, P.~M., \& Santos-Lle{\' o}, M.\ 2004, \mnras, 351, 161 

\bibitem[Piconcelli et al.(2005)]{2005A&A...432...15P}
Piconcelli, E., Jimenez-Bail{\' o}n, E., Guainazzi, M., Schartel, N., Rodr{\'{\i}}guez-Pascual, P.~M., \& Santos-Lle{\' o}, M.\ 2005, \aap, 432, 15 

\bibitem[Porquet et al.(2004)]{2004A&A...422...85P}
Porquet, D., Reeves, J.~N., O'Brien, P., \& Brinkmann, W.\ 2004, \aap, 422, 85 

\bibitem[Pounds et al.(1990)]{1990Natur.344..132P}
Pounds, K.~A., Nandra, K., Stewart, G.~C., George, I.~M., \& Fabian, A.~C.\ 1990, \nat, 344, 132 

\bibitem[Pounds et al.(2003)]{2003MNRAS.345..705P}
Pounds, K.~A., Reeves, J.~N., King, A.~R., Page, K.~L., O'Brien, P.~T., \& Turner, M.~J.~L.\ 2003a, \mnras, 345, 705 

\bibitem[Pounds et al.(2003)]{2003MNRAS.346.1025P}
Pounds, K.~A., King, A.~R., Page, K.~L., \& O'Brien, P.~T.\ 2003b, \mnras, 346, 1025 

\bibitem[Pounds et al.(2004)]{2004ApJ...616..696P}
Pounds, K.~A., Reeves, J.~N., Page, K.~L., \& O'Brien, P.~T.\ 2004, \apj, 616, 696 

\bibitem[Ross \& Fabian(1993)]{1993MNRAS.261...74R}
Ross, R.~R., \& Fabian, A.~C.\ 1993, \mnras, 261, 74 

\bibitem[R{\' o}{\. z}a{\' n}ska \& Czerny(2000)]{2000MNRAS.316..473R}
R{\' o}{\. z}a{\' n}ska, A., \& Czerny, B.\ 2000, \mnras, 316, 473 

\bibitem[R{\' o}{\. z}a{\' n}ska et al.(2002)]{2002MNRAS.332..799R}
R{\' o}{\. z}a{\' n}ska, A., Dumont, A.-M., Czerny, B., \& Collin, S.\ 2002, \mnras, 332, 799 

\bibitem{rozanska2004}
R\' o\. za\' nska, A., Czerny B., Siemiginowska A., Dumont A.-M., \& Kawaguchi T., 2004, ApJ, 600, 96

\bibitem{rozanska2005}
R\' o\. za\' nska, A., Goosmann, R.W., Dumont, A.-M., Czerny, B. 2005, preprint

\bibitem{sako2003}
Sako M. et al. 2003, ApJ, 596, 114

\bibitem[Sambruna et al.(2001)]{2001ApJ...546L..13S}
Sambruna, R.~M., Netzer, H., Kaspi, S., Brandt, W.~N., Chartas, G., Garmire, G.~P., Nousek, J.~A., \& Weaver, K.~A.\ 2001, \apjl, 546, L13 

\bibitem{sobolewska2005}
Sobolewska, M., \& Done, C. 2005, Proceedings for XDAP 2004, to be published in AIP

\bibitem[Steenbrugge et al.(2005)]{2005A&A...432..453S}
Steenbrugge, K.~C., Kaastra, J.~S., Sako, M., Branduardi-Raymont, G., Behar, E., Paerels, F.~B.~S., Blustin, A.~J., \& Kahn, S.~M.\ 2005, \aap, 432, 453 

\bibitem[Tanaka et al.(1995)]{1995Natur.375..659T}
Tanaka, Y., et al.\ 1995, \nat, 375, 659 

\bibitem[Turner et al.(2003)]{2003MNRAS.346..833T}
Turner, A.~K., Fabian, A.~C., Vaughan, S., \& Lee, J.~C.\ 2003, \mnras, 346, 833

\bibitem[Wilkes \& Elvis(1987)]{1987ApJ...323..243W}
Wilkes, B.~J., \& Elvis, M.\ 1987, \apj, 323, 243 

\bibitem{young2005}
Young, A.J., Lee, J.C., Fabian, A.C., Reynolds, C.S., Gibson, R.R., \& Canizares, C.R. 2005, astro-ph/0506082

\end{thebibliography}
\end{document}